\documentclass[twocolumn,floats,prb]{revtex4}

\usepackage{amsmath}
\usepackage{amsfonts}
\usepackage{amssymb}
\usepackage{graphicx}
\usepackage{hyperref}

\newcommand{\beq}{\begin{equation}}
\newcommand{\eeq}{\end{equation}}
\newcommand{\bea}{\begin{eqnarray}}
\newcommand{\eea}{\end{eqnarray}}
\newcommand{\beano}{\begin{eqnarray*}}
\newcommand{\eeano}{\end{eqnarray*}}
\newcommand{\no}{\nonumber}

\newcommand{\ra}{\rangle}
\newcommand{\la}{\langle}

\newcommand{\bb}{{\bar b}}
\newcommand{\fb}{{\bar f}}
\newcommand{\bd}{b^\dagger}
\newcommand{\fd}{f^\dagger}
\newcommand{\bph}{b^{\vphantom{\dagger}}}
\newcommand{\fph}{f^{\vphantom{\dagger}}}
\newcommand{\bbd}{\bb^\dagger}
\newcommand{\fbd}{\fb^\dagger}
\newcommand{\fbph}{{\bar f}^{\vphantom{\dagger}}}
\newcommand{\bbph}{{\bar b}^{\vphantom{\dagger}}}

\newcommand{\Rb}{\bar{R}}

\newcommand{\st}[2]{\bigl|#1,#2\bigr\ra}
\newcommand{\bst}[2]{\bigl| \overline{#1,#2}\bigr\ra}
\newcommand{\sst}[1]{|#1\ra}
\newcommand{\bsst}[1]{|\overline{#1}\ra}
\newcommand{\bC}{\bar{C}}
\newcommand{\bD}{\bar{D}}
\newcommand{\bX}{\bar{X}}
\newcommand{\bY}{\bar{Y}}
\newcommand{\bR}{\bar{R}}

\newcommand{\lie}{\text{Lie}}

\begin{document}

\title{Localization in disordered superconducting wires with broken
spin-rotation symmetry}

\author{Ilya A. Gruzberg$^1$, N. Read$^2$, and Smitha Vishveshwara$^3$}

\affiliation{$^1$The James Franck Institute, The University of Chicago,
5640 S. Ellis Avenue, Chicago, IL 60637 \\
$^2$Department of Physics, Yale University, P.O. Box 208120, New
Haven, CT 06520-8120 \\
$^3$Department of Physics, University of Illinois at Urbana-Champaign,
Urbana, IL 61801}

\date{December 15, 2004}
%\date{\today}

\begin{abstract}
Localization and delocalization of non-interacting quasiparticle
states in a superconducting wire are reconsidered, for the cases
in which spin-rotation symmetry is absent, and time-reversal
symmetry is either broken or unbroken; these are referred to as
symmetry classes BD and DIII, respectively. We show that, if a
continuum limit is taken to obtain a Fokker-Planck (FP) equation
for the transfer matrix, as in some previous work, then when there
are more than two scattering channels, all terms that break a
certain symmetry are lost. It was already known that the resulting
FP equation exhibits critical behavior. The additional symmetry is
not required by the definition of the symmetry classes; terms that
break it arise from non-Gaussian probability distributions, and
may be kept in a generalized FP equation. We show that they lead
to localization in a long wire. When the wire has more than two
scattering channels, these terms are irrelevant at the short
distance (diffusive or ballistic) fixed point, but as they are
relevant at the long-distance critical fixed point, they are
termed {\em dangerously irrelevant}. We confirm the results in a
supersymmetry approach for class BD, where the additional terms
correspond to jumps between the two components of the sigma model
target space. We consider the effect of random $\pi$ fluxes, which
prevent the system localizing. We show that in one dimension the
transitions in these two symmetry classes, and also those in the
three chiral symmetry classes, all lie in the same universality
class.

\end{abstract}

\pacs{PACS numbers: }

\maketitle

\section{Introduction}

Localization of single particle states in disordered fermionic
systems has been a key concept underlying our understanding of
disordered normal metals. The nature of the states depends
crucially on the presence or absence of time reversal (TR) and
spin rotation (SR) symmetries. Based on these considerations,
three distinct Wigner-Dyson symmetry classes were identified a
long time ago --- orthogonal, unitary and symplectic (the names
come from the random matrix theory of Wigner and Dyson
\cite{Mehta}).

Recently, it has been realized that localization ideas are useful
in describing a wider class of systems including superconductors
and superfluids, various quantum Hall states and random-bond Ising
models. A crucial part of the emerging picture is again the role
played by symmetries of the disordered single-particle
Hamiltonians describing these systems. In addition to the TR and
SR symmetries, these new systems may possess additional symmetries
that have to be taken into account. One such category is comprised
of systems defined on bipartite lattices with the sublattice (also
called chiral) symmetry. Another category, on which this paper is
focussed, includes Bogoliubov-de Gennes (BdG) Hamiltonians which
describe quasiparticles in disordered superconductors at a mean
field level. Although particle number is not conserved in the mean
field description, the problem of finding the quasiparticle
spectrum can be transformed to a particle-number conserving form,
so that it reduces to solving a single-particle Hamiltonian that
has certain symmetries. The spectra of Hamiltonians in all of
these classes have an exact reflection symmetry around zero
energy, and eigenstates near zero energy may have localization
properties different from those of eigenstates in other parts of
the spectra.

A symmetry classification of disordered single-particle
Hamiltonians has been established by Altland and Zirnbauer (AZ)
\cite{AZ,Z}. The symmetry class and the dimensionality of a
particular disordered system in principle determine which phases
(localized or extended) and phase transitions may occur in its
phase diagram. After ensembles of random matrices, which are the
zero-dimensional cases, the next simplest situation arises in one
dimension (1D), where various analytical methods are available. In
the standard Wigner-Dyson classes all the single-particle states
are localized in 1D and this leads to an exponential decrease of
the conductance of a 1D wire with its length, that is, the
so-called strong localization.

Brouwer {\it et al.} in Ref.~\onlinecite{BFGM} studied
localization in 1D (superconducting wires) using the Fokker-Planck
(FP) scattering approach \cite{Dorokhov,MPK}. They considered all
four BdG symmetry classes (denoted C, CI, D, and DIII in
Ref.~\onlinecite{AZ}), and obtained the corresponding four FP
equations describing transport at zero energy. While for the two
classes C and CI that have unbroken SR invariance the
quasiparticle (thermal) conductance $\cal G$ decays exponentially
with the length $L$ of the wire for large $L$, the situation was
found to be quite different in classes D and DIII in which the SR
invariance is absent. In the latter, the mean $\langle {\cal G}
\rangle$ decays only algebraically to zero for large $L$ and $\ln
{\cal G}$ is not self-averaging, indicating a very broad
distribution of the conductance and the absence of the exponential
localization of the quasiparticle states. This behavior was termed
critical in Ref.~\onlinecite{BFGM}.

This surprising result was questioned by Motrunich {\it et al.} in
Ref.~\onlinecite{Motrunich}. In this paper some models in symmetry
classes D and DIII were studied using real-space renormalization
group methods, and it was argued that, for a generic distribution
of disorder, at zero energy there are localized phases separated
by a phase boundary at which the localization length diverges. It
was found that close to the critical point, the low-energy density
of states (DOS) per unit length, $\nu(\epsilon)$, has a divergent
power-law behavior with a non-universal varying exponent:
\begin{align}
\nu(\epsilon) \propto \epsilon^{-\delta}, \label{DOS-varying}
\end{align}
as $\epsilon\to0^+$, with $\delta<1$. At the critical point, one
has $\delta = 1$, and there are logarithmic corrections of the
Dyson form \cite{Dyson}:
\begin{align}
\nu(\epsilon) \propto \epsilon^{-1}|\ln\epsilon|^{-3}.
\label{DOS-Dyson}
\end{align} In
addition, as parameters vary, it is possible for the phase
boundaries to collide, in which case there is a multicritical
point with different critical properties.

In Ref.~\onlinecite{TBFM}, the FP scattering approach was used to
study the DOS in disordered wires in all symmetry classes. Perhaps
unsurprisingly, for the wires in classes D and DIII the result had
a singularity of the critical form in Eq.~(\ref{DOS-Dyson}). In
Ref.~\onlinecite{BFM}, Brouwer {\it et al.} tried to account for
the discrepancies between the results of the two groups in terms
of a crossover from the unitary class to class D within a
particular model. We believe that this account leaves the role of
symmetries and universality in the problem unclear, and that our
work described here greatly clarifies these questions.

In this paper we reconsider localization in superconducting wires
with broken spin-rotation symmetry, in classes BD and DIII in the
AZ classification [we prefer the term BD in place of class D for
more than zero dimensions, as the distinction between odd (B) and
even (D) size Hamiltonian matrices disappears for infinite
matrices], for a finite number $N$ of propagating modes in the
wires, and resolve the major issues. The central point concerns
symmetries of the problem. The most popular type of models for
disordered wires are those in which propagation along the wire is
continuous, and this applies in particular for weak disorder. We
show that for the cases of classes BD and DIII, there are certain
symmetry operations on the spaces of transfer matrices (or on the
probability distributions for the disorder) which have been
overlooked previously. The conditions that define classes BD and
DIII do {\em not} impose that the distribution be invariant under
these operations. However, when one takes the standard limit that
results in the FP equation, the leading symmetry-breaking term is
a differential operator of order $N$, and terms of order greater
than two disappear in the limit (these are related to deviations
of the disorder distribution from Gaussian). That is, {\em the FP
equations for classes BD and DIII studied by Brouwer {\it et al}
\cite{BFGM,TBFM} possess more symmetry than the general class of
distributions does}. For $N=1$, the leading symmetry-breaking term
is a drift (first-order) term which produces localization if it is
present. We show that for a long wire, the $N$ channel case
reduces to the $N=1$ situation, and the leading symmetry-breaking
term reduces to the drift term. Then generically, {\em breaking
the additional symmetry produces localization}. In a
generalization of the FP equation that keeps higher-order
operators, operators of order higher than 2 would be irrelevant at
short distance [in renormalization group (RG) language]. Since
they are nonetheless relevant at large distances (they take the
system off criticality), they cannot be dropped, and so they are
analogous to ``dangerously irrelevant'' operators in critical
phenomena. In practice, to represent a system that will have both
a finite number $N$ of channels, and a finite mean free path as an
ultraviolet cutoff, the FP limit is not taken, and
symmetry-breaking terms must be retained. The behavior near the
critical point is universal, and is governed by the $N=1$ case.
The results agree with those of Motrunich {\it et al\/}
\cite{Motrunich}.

In addition to demonstrating these points with the FP approach, we
also give the supersymmetric representation of the wire for class
BD, and show how the phenomena appear there. The wire is
represented by a finite-$N$ analog of the supersymmetric nonlinear
sigma model (in 1D), which is known to have a target space that
consists of two connected components in the cases of classes BD
and DIII \cite{BSZ}. We show that the terms in the distribution of
disorder that break the additional symmetry produce a jump from
one component to the other whenever they occur in the expansion of
a disorder average. These are domain walls in the sigma model;
topologically, in 1D these domain walls are simply points. The
supersymmetry approach is complementary to the FP equation; it
produces results for the moments of Green's functions, and is used
here to study the DOS.

For the more general situation in which the transfer matrix does
not have to evolve continuously, as in a lattice (or network)
model, it is possible for an amplitude to pick up a factor of $-1$
during a time step, which would correspond to encircling a flux of
$\pi$ in a lattice model; such an effect is consistent with the
conditions that define classes BD and DIII. It was already shown
\cite{Chalk} in the case of class BD with $N=1$ that if such
events are distributed independently and uniformly along the wire,
the drift term in the FP equation is effectively suppressed, and
the system is again critical. We show here that this also occurs
for the case of general $N$, and then on scales larger than the
separation of these events, the $N$-channel problem effectively
maps onto the analysis as given in Ref.~\onlinecite{BFGM}, and is
always critical even when the symmetry-breaking terms are present.
In the supersymmetry approach, domain walls are suppressed in this
situation. Thus domain walls are essential to produce localization
for all $N$, as is found also in two dimensions for class BD
\cite{BSZ,Chalk}.

It is striking that the universal results for the symmetry classes
BD and DIII turn out to be the same, and also are the same as
those for the chiral classes. We show that this
``super-universality'' can be understood, as the $N=1$ channel
cases all possess hopping Hamiltonian models that can be mapped
onto each other exactly.

The rest of the paper is structured as follows. In Section
\ref{sec:symmetry} we provide a brief account of the scattering
approach to coherent transport in wires, stressing its symmetry
aspects, and including the role of non-Gaussian distributions,
especially in the case of classes BD and DIII. In Section
\ref{sec:GenFP} we review the standard FP equation resulting from
this approach, and its solution. We generalize the FP equation to
include the leading terms that break the extra symmetries. We
simplify the resulting generalized FP equation in the asymptotic
regime of long wires and solve it, obtaining the scaling function
of the logarithm of the conductance near the critical point, and
exhibiting two localization lengths, the typical localization
length for the typical values (and moments) of the logarithm of
the conductance, and the mean localization length for the moments
of the conductance itself. In Section \ref{sec:SUSY}, we describe
aspects of the supersymmetry method for wires in the BD class.
This gives us the scaling of the mean localization length and the
mean density of states near the critical point. In Section
\ref{sec:flux} we explain what occurs in more general models in
which the propagation along the wire is discontinuous, allowing
fluxes of $\pi$ to be randomly inserted. The mappings that
establish super-universality are given in Section
\ref{sec:superu}. Some final discussion of universality is given
in the conclusion, Section \ref{sec:discussion}. Finally, many
technical details are presented in Appendices.

\section{Symmetry aspects of the transfer matrix}
\label{sec:symmetry}

In this section we consider the symmetry aspects within the
transfer matrix formulation, which will be used in both the FP and
supersymmetry approaches, and show that for classes BD and DIII
with more than two scattering channels there are additional
symmetries present in a Gaussian distribution of disorder that are
not generically there. In order to make the key points in our
analysis, it will be useful to reconsider briefly the basics of
the transfer matrix approach to one-dimensional (1D) quantum
transport (for more details, see Ref.~\onlinecite{Beenakker}).

\begin{figure}%[t]
\includegraphics*[width=3.4in]{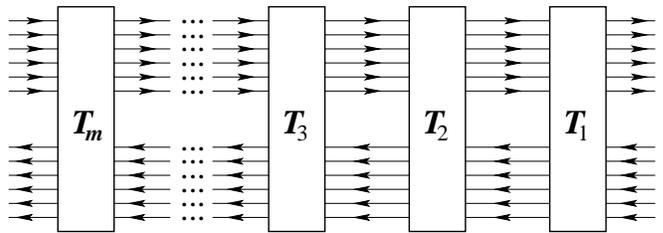}
\caption{The transfer matrix for the wire is composed of $m$
steps, with independently-distributed increments $T_j$, $j=1$,
\ldots, $m$.} \label{fig:wire}
\end{figure}

It is convenient to pass from the single-particle Hamiltonian
$\cal H$ to a scattering description, by considering waves of a
given energy that enter the system at either end, are scattered
and then emerge at the same or the opposite end. We will assume
that only a finite number of ``channels'' are relevant at the
energy considered. For the non-standard ensembles, the energy
chosen has the special value $\epsilon = 0$ (this corresponds to
calculating the transport properties of the ground state of the
superconductor). The scattering can then be represented by an
$S$-matrix that maps incoming to outgoing waves, or, better, by a
transfer matrix $M'$ that maps the in- and outgoing waves at one
end (say, the right) to those at the other (the left); both of
these matrices are $2N'\times 2N'$.  The advantage of the transfer
matrix is that 1D systems can be composed end-to-end by joining
together short elements, and multiplying the respective transfer
matrices, see Fig. \ref{fig:wire}. Taking the latter as random,
identically-distributed, statistically-independent matrices, leads
to an equation that describes the evolution of the probability
distribution for the transfer matrix with increasing length of the
system.

The transfer matrix is related to the Landauer conductance of the
system. The Landauer-Buttiker formula (see
Ref.~\onlinecite{Stone+Szafer} for history) for the two-probe
(electrical) conductance reads
\begin{align}
{\cal G} = \text{tr } \boldsymbol{t}\boldsymbol{t}^{\dagger},
\end{align}
where $\boldsymbol{t}$ is the $N'\times N'$ transmission matrix
for waves entering at one end to exit at the other. In
superconductors, charge (or quasiparticle number) is not a good
quantum number at mean field level, but the Landauer conductance
is still relevant for the thermal conductance, since heat is
transported by quasiparticles: the thermal conductance is given by
$\frac{\pi^2}{6} k_B^2 T {\cal G}$. The conductance can be
rewritten in terms of the transfer matrix as
\begin{align}
{\cal G} = 2 \, \text{tr} \, \bigl(2 + {M'}{M'}^\dagger +
({M'}{M'}^\dagger)^{-1}\bigr)^{-1}. \label{conductance2}
\end{align}
Thus it is a trace of a function only of ${M'}{M'}^\dagger$ (or of
$M'^\dagger M'$).

The mathematical properties of $M'$ are as follows. Because of
conservation of probability flux (``current'') in time-independent
scattering, the $S$-matrix is unitary, which implies that the
transfer matrix is pseudounitary:
\begin{align}
{M'}^\dagger \Lambda {M'} = \Lambda, \label{}
\end{align}
where $ \Lambda = \Bigl(
\begin{array}{cc}
I_{N'} & 0 \\
0 & -I_{N'}
\end{array}
\Bigr) $. [The first $N'$ indices refer to waves entering at the
right (or leaving at the left), the second to those leaving at the
right (or entering at the left).] Thus $M'$ is a member of the
noncompact Lie group U($N',N'$). For the unitary ensemble, this is
the only condition, but for the other ensembles \cite{AZ},
additional conditions are imposed on $M'$. They result from the
symmetries of the Hamiltonian $\cal H$ that define the symmetry
class, and have the effect of restricting $M'$ to a noncompact
subgroup $G$ of U($N',N'$). In some cases, the representation of
$G$ in U($N',N'$) may be reducible \cite{BFGM}. That is, there
exists a unitary transformation to a basis in which the full $2N'
\times 2N'$ transfer matrix $M'$ becomes block diagonal, each of
the $d$ blocks being (for symmetry reasons) the same $2N \times
2N$ transfer matrix $M$ (or in some cases, its adjoint, or
transpose), so $N'=dN$, where $d$ depends on the symmetry class.
We will call $N$ (rather than $dN$) the number of channels, as it
is the number of independent variables propagating in each
direction.

In this section, and in the remainder of this paper except Section
\ref{sec:flux}, we aim to model the situation in which the
propagation along the wire is continuous. For calculational
convenience, we nonetheless represent this by discrete steps, as
multiplication of transfer matrices for each step. The underlying
continuity implies that after each step, $M'$ lies in $G$ and can
be continuously connected to the identity, and thus $G$ must be a
connected group. For class BD, $M'=M$ can be taken as real, so $
G$ is SO$_0$($N,N$) with $N'=N$ (as always, the prefix S means
unit determinant, while the subscript $0$ denotes the component
containing the identity). For class DIII, $G$ is isomorphic to
SO($2N,\mathbb{C}$) ($N'=2N$). {\em For the purposes of most of
this paper, these statements can be taken as the definitions of
these symmetry classes.} Tight-binding models of superconductors,
though maybe not strictly continuous, lead to these systems with
$N$ even in both symmetry classes: Pairing of spinless fermions
(without time-reversal symmetry) gives class BD, while pairing of
spin-$1/2$ fermions with time-reversal symmetry, but with broken
spin-rotation symmetry, gives class DIII. Pairing of fermions with
spin $1/2$, but with spin-rotation and time-reversal symmetries
broken, gives class BD with $N$ divisible by 4. Our conventions
differ from those of Ref.~\onlinecite{BFGM}, where their $N$ is
half ours for class DIII, and one-quarter of ours for class BD
because they considered the case of spin-1/2 fermions. Similarly,
some of the models studied in Ref.~\onlinecite{Motrunich} are in
class BD with $N=2$ or $4$, and others are in class DIII with
$N=2$. Our analysis applies to all positive integer values of $N$
in both classes, and one odd value, $N=1$, will play a special
role in the analysis.

Mathematically, class DIII Hamiltonians for a 1D wire can also
lead to transfer matrices in SO$(2N+1,\mathbb{C})$. Here we would
have $N'=2N+1$ instead of $N'=2N$. The behavior of these systems
is different from that of those we analyze here, and somewhat
similar to the odd number of channels case of the symplectic (AII)
symmetry class \cite{takane}; namely the conductance of a long
wire approaches ${\cal G}=1$, an effect that was found in
Ref.~\onlinecite{Zsymp}. These ``$N+1/2$-channel'' class DIII
systems can be realized as edge states of certain class DIII
superconductors in two space dimensions. We do not consider these
cases further in this paper.

The formula (\ref{conductance2}) for the conductance becomes, in
terms of $M$,
\begin{align}
{\cal G} = 2d \, \text{tr} \, \bigl(2 + {M}{M}^\dagger +
({M}{M}^\dagger)^{-1}\bigr)^{-1}, \label{conductance3}
\end{align}
and is invariant under multiplication of $M$ on the left {\it
or\/} right by unitary matrices. Those that lie in $G$ represent
some redundant aspect of the evolution of $M$, in the sense that
some scattering does not affect the conductance. We can find a
maximal unitary (hence compact) subgroup $K$ of $G$, such that
$\cal G$ is invariant under ${M} \to k_1 {M} k_2, \quad k_1,k_2
\in K$. This suggests parameterizing $M$ so as to factor off this
subgroup. For example, for class BD, $K = \text{SO}(N)\times
\text{SO}(N)$, and for DIII, $K = \text{SO}(2N)$. Then any
transfer matrix $M$ for class BD can be written schematically as
$M=k_1ak_2$, or explicitly
\begin{align}
{M} = \begin{pmatrix} V_1 & 0 \\ 0 & V_2 \end{pmatrix}
\begin{pmatrix} \cosh X & \sinh X \\ \sinh X & \cosh X \end{pmatrix}
\begin{pmatrix} V_3 & 0 \\ 0 & V_4 \end{pmatrix},
\label{Cartan-decompo-groupbd}
\end{align}
where $V_1,\ldots,V_4 \in \text{SO}(N)$. For class DIII, we have
instead
\begin{align}
{M} =  V_1
\begin{pmatrix} \cosh X & i\sinh X \\ -i\sinh X & \cosh X
\end{pmatrix}
V_2,  \label{Cartan-decompo-groupdiii}
\end{align}
and $V_1$, $V_2$ are in SO($2N$). Thus the group can be decomposed
as $G=KAK$ (the Cartan decomposition of $G$), where $A$ stands for
the matrices containing $X$ in the above expressions. In both
cases, $X = \text{diag}(x_1,\ldots,x_N)$ is real and diagonal.
These parametrizations are also called radial or polar
decompositions. The real parameters $x_i$ are called radial
coordinates, and the matrices $V_1$, \ldots, $V_4$, (resp., $V_1$,
$V_2$) represent angular coordinates. Similar decompositions exist
for other groups. A familiar example of the Cartan decomposition
in a compact group is to take $G=$ SU($2$), $K=$ U($1$) (generated
by the $\sigma_z$ Pauli matrix), and then the three parameters in
the decomposition are exactly the Euler angles that describe a
rotation.

If we identify elements of $G$ modulo right multiplication by
elements of $K$, we obtain the coset space $G/K$. If we also
identify points of this space modulo the action of $K$ given by
left multiplication, then we obtain the double coset space
$K\backslash G/K$. The $x_i$ parametrize this double coset space
$K\backslash G/K$, and the Landauer conductance $\cal G$ is a
function only of these radial coordinates:
\begin{align}
{\cal G} = d\sum_{i=1}^N \frac{1}{\cosh^2 x_i}.
\label{Landauer-conductance}
\end{align}
To ensure that each point of $K\backslash G/K$ is labeled by a
single set of $x_i$, there should be restrictions on the possible
values of $x_i$. We notice that it is possible to map the set $A$
into itself by conjugation by certain elements of $K$, $a\to
kak^{-1}$. Since left and right multiplication by $K$ is to be
divided out, these give the same point in the double coset space.
The group of such operations is called the Weyl group of the
Cartan decomposition in question. By inspecting $K$ for the cases
of class BD and DIII, we see that it is possible in this way to
permute the $x_i$s, and to reverse the sign of an {\em even}
number of $x_i$s [because for class BD, $K$ involves SO($N$), not
O($N$), and similarly for DIII]. Consequently, a unique
parametrization for $K\backslash G/K$ is obtained by using the
following domain (known as the Weyl chamber) for $x_i$:
\begin{align}
C &= \{|x_1| < x_2 < \ldots < x_N\}, \label{Weyl-chamber}
\end{align}
where $x_1$ can be either positive or negative.

We want to note here that all the previous treatments of classes
BD and DIII in 1D based on the scattering approach, including
Refs.~\onlinecite{BFGM,TBFM,BFM}, incorrectly assumed a different
range for $x_i$:
\begin{align}
0 < x_1 < x_2 < \ldots < x_N.
\label{Weyl-chamber-1}
\end{align}
This range is the one appropriate for all the standard
(Wigner-Dyson) symmetry classes, and for the BdG classes C and CI
in the AZ classification. For the chiral classes, the range is
\begin{align}
x_1 < x_2 < \ldots < x_N.
\label{Weyl-chamber-chiral}
\end{align}
The property that at least one of the radial coordinates ($x_1$)
can take negative values is shared by the classes BD, DIII, and
the chiral ones. Ultimately, as we show below, this allows phase
transitions to occur in these classes. Within the FP approach, a
transition at which the localization length diverges occurs when
the mean of an $x_i$ (in the limit of a long wire) passes through
zero. There is one such transition in classes BD and DIII, but $N$
in the chiral classes. (Related distinctions between the two
phases, implying that there must be a transition, appeared in
Refs.~\onlinecite{Motrunich,merz}.) The distinction in the ranges
for $x_i$ is crucial to our symmetry analysis also, as we will see
that the additional symmetry mentioned in the introduction acts on
the double coset space by reversing the sign of $x_1$.

We next consider $M$ as a product of $m= L/a$ elements of $G$:
\begin{align}
{M}_m = T_m T_{m-1} \ldots T_1. \label{prodT}
\end{align}
This discrete form could represent a discrete 1D model with
lattice spacing $a$, or could be used to take a continuum limit $a
\to 0$ with $L$ fixed. In order to increase the length from $m$ to
$m+1$, we multiply ${M}_m$ on the left by $T_{m+1}$. In general,
we take $T_j$ to be a random variable, in which case we obtain a
stochastic process. If we assume that $T_j$s at different $j$ are
statistically independent, and that all $T_j$ are identically
distributed, then we obtain a process whose dynamics is
``time''-independent, where by ``time'' we mean position $j$ along
the wire (this terminology will be used consistently throughout
this paper). Also, the time-evolution process is invariant under
right multiplication by any element of the full group $G$, in the
sense that $(T_{m+1}M)g = T_{m+1}(Mg)$ for $g \in G$, and in
particular under right multiplication by any element of $K$. Thus
it can be viewed as a stochastic process on the coset space $G/K$
(and this is the point of view most often taken in the literature,
see Refs.~\onlinecite{Huffmann}, \onlinecite{Caselle}). But it is
not in general invariant under left multiplication even by
elements of $K$, unless the distribution of $T_j$ is invariant
under $T_j \to k T_j k^{-1}$ for all $k \in K$ and all $j$. Unless
this assumption is made, we do not obtain a well-defined
stationary stochastic process on the double coset space
$K\backslash G/K$, because the effect of left-multiplication by
$T_{m+1}$ on a double coset $KgK$ (where each $K$ stands for all
members of $K$) depends on which element of the $K$ is chosen at
the left.

It is generally believed, however, that even in a generic
situation, due to the compactness of the subgroup $K$, the
diffusion in the angular variables will lead eventually (for long
wires) to a uniform distribution on $K$ (an ergodicity
assumption). Therefore, one can use $K$-invariant distributions
for $T_j$ to analyze localization properties in long wires,
without loss of generality. Together with the $K$-invariant
initial condition $M=I$ for zero length, we can view the process
as taking place on the double coset $K\backslash G/K$. All
information necessary to obtain the probability distribution for
$\cal G$ is contained in the joint distribution function of the
radial coordinates $P(x_1,...x_N;L) \equiv P(x;L)$. It depends on
the length of the wire $L$, and our goal is to find the equation
governing the evolution of $P(x;L)$ when $L$ increases.

Each $T_{m+1}$ is the transfer matrix of a thin slice, and as such
is best parameterized as $ T_{m+1} = \exp A $. Here $A$ belongs to
the Lie algebra of $G$, $\lie(G)$, and the probability
distribution for $A$ is assumed to be concentrated near zero since
the slice is thin. Also, due to the invariance of the
distribution, $A$ may be assumed to be in the subspace $\cal{P}$
of $\lie(G)$ which appears in the infinitesimal version of the
Cartan decomposition $ \lie(G) = \lie(K) + \cal{P}$. (An element
of $\cal P$ can be diagonalized using elements of $\lie(K)$, as we
used earlier, but we do not do this here.) Let us illustrate this
point for class BD. In this case $\lie(G) = \text{so}(N,N)$,
$\lie(K) = \text{so}(N) + \text{so}(N)$, and the subspace
$\cal{P}$ consists of matrices of the form
\begin{align}
A = \begin{pmatrix} 0 & \theta \\ \theta^T & 0 \end{pmatrix},
\label{theta}
\end{align}
where $\theta$ is a real $N\times N$ matrix. This form for $A$
corresponds to motion in $G/K$ starting from the point $M_mK$, and
relative to such an origin the matrices $\theta$ parametrize
$G/K$. For class DIII, $A=\theta$ is a purely imaginary
antisymmetric $2N\times 2N$ matrix.

It will be useful to examine the case $N=1$ for class BD
explicitly as motivation for the symmetry arguments that follow,
and because it plays a central role in the analysis even for
general $N$. In this case, $G=$ SO$_0(1,1)$, $K$ is trivial, and
$M$ is described by a single real number $x_1$, which is easily
seen to be $x_1=\sum_j \theta_j$. Thus the evolution process is a
random walk on the real line parametrized by $x_1$ or $\theta$.
Clearly, if this random walk has net drift of either sign (i.e.\
the mean of $\theta_j$ is nonzero), then $x_1$ goes to plus or
minus infinity with high probability, and the conductance becomes
exponentially small, indicating localization. If the mean of
$\theta_j$ is zero, the distribution of $x_1$ will have width of
order $\sqrt{L}$ (we assume that the distribution for $\theta_j$
has finite variance). In this case, the behavior is of the same
form as the critical behavior for class BD, mentioned in the
introduction \cite{BFGM}. This behavior will be considered in more
detail below; here we wish only to draw attention to the central
feature, which is that there is an obvious symmetry operation on
SO$_0(1,1)$, given by reversing the sign of $x_1$, and that the
drift breaks this symmetry. This symmetry is a discrete operation,
and cannot be obtained by using the left and right action of
SO$_0(1,1)$ on itself. There is a similar analysis for $N=1$ in
class DIII: SO($2,\mathbb{C}$) can be represented by the group of
nonzero complex numbers $z$ under multiplication (in fact, the
eigenvalues of $M$ are $z$, $z^{-1}$), $K=$ SO(2) is represented
by the complex numbers of modulus 1, and the additional symmetry
is given by $z \to z^{-1}$. Localization occurs if $z$ drifts to
either $0$ or $\infty$; here $x_1=\ln |z|$.

As an aside, we see from this analysis that the double coset
spaces for classes BD and DIII for $N=1$ are the same. In fact,
the symmetric spaces $G/K$ for the two cases are also the same
space: both are simply the real line, with translation symmetry
along the line (in suitable variables). For DIII this relies on
the fact that SO$(2,\mathbb{C})$ is Abelian, and so $K\backslash
G/K$ and $G/K$ are the same. Moreover, there is a similar symmetry
discussion for all three chiral symmetry classes (orthogonal,
unitary, and symplectic, or classes BDI, AIII, and CII
respectively in the AZ classification), and for these the double
coset and symmetric spaces are also the same space (the real line)
for $N=1$. Therefore, at least for properties that can be
expressed in terms of the transfer matrix, the critical behavior
that occurs in each of these cases is the same. While this
observation may seem trivial at this stage, it will become much
more significant once we see that the $N=1$ cases control the
long-distance behavior of the critical systems at $N>1$ also. Such
equivalences, which rely on well-known isomorphisms of Lie groups
of small rank, also occur in some other cases of the transfer
matrix approach to disordered wires. In Section \ref{sec:superu}
below, we establish mappings between the $N=1$ cases in the above
five classes directly, in terms of hopping Hamiltonians.

It is natural to ask whether there are any analogous additional
symmetries for classes BD and DIII in the cases with $N>1$. It
turns out that there are. In both cases, there is a larger group
that contains $G$ as a connected subgroup: these are O($N,N$) and
O($2N,\mathbb{C}$). O($N,N$) has four connected components,
including two in which the elements have determinant $-1$, and the
part SO($N,N$) with determinant $+1$ has two connected components,
one of which is SO$_0(N,N)$. O$(2N,\mathbb{C})$ has just two
components, with determinants $\pm 1$ respectively. Certain
subgroups of these act on our groups $G$, mapping it into itself.
In particular, as we wish to leave the identity element fixed,
there is the action by conjugation, $g\to hgh^{-1}$, where $g\in
G$, and $h$ is in the larger group. Examples for $h$ of the
desired form can in fact be found by choosing $h$ in
O$(N)\times$O$(N)$ or O$(2N)$, for class BD or DIII respectively;
in both cases, such an $h$ has the desired effect if it has
negative determinant. Then the product of any two such $h$ lies in
the subgroup S[O$(N)\times$O$(N)$] or SO($2N$), respectively. (In
the special case of class BD with $N$ odd, we can take $h$ to be
$\Lambda$, which commutes with any element of $K$.) For the action
on the double coset space, we can compare this discussion with
that of the Weyl group, and we see that we now have the set of all
permutations and sign reversals of the $x_i$, so that in terms of
the Weyl chamber $C$, the additional symmetry is simply to reverse
the sign of $x_1$. Mathematically, what we have described is an
outer automorphism (or involution) of the reduced root system of
$G/K$ (outer means it is not connected to the identity). As these
are of type $D_N$ in both cases, and the outer automorphisms are
known to be related to the existence of symmetries of the Dynkin
diagram, we do find such a symmetry. [There are similar additional
symmetries for each of the chiral symmetry classes. In these
cases, the operation that sends $x_i$ to $-x_i$ for {\em all} $i$
is such a symmetry, an outer automorphism of the reduced root
systems, which are all of type $A_{N-1}$ for these classes.
However, these symmetries cannot be written as conjugation by
elements of a covering group of $G$, unlike the BdG cases above.]

We saw in the examples with $N=1$ that if the distribution of the
random transfer matrices $T_j$ for each slice is invariant under
the additional symmetry (as well as under conjugation by $K$),
then localization would be prevented. As localization requires
that all $x_i$ including $x_1$ go off to $\pm \infty$, it is now
clear that the same will be true for $N>1$. On the other hand, if
the distribution is perturbed so as to break the symmetry, then
localization may set in. [For the chiral classes, there is a
similar effect: if the symmetry under $x_i\to -x_i$ for all $i$ is
preserved, then localization does not occur (the system is
critical) if and only if $N$ is odd\cite{BMSA}.] We must now
determine the form of probability distributions that satisfy the
symmetry requirements.

A popular, and usually effective, choice is to take the
distribution for the matrix $\theta$ in each time slice to be
Gaussian,
\begin{align}
P[A] \propto \exp\left( -\frac{C_2\ell}{a} {\rm tr}\,A ^2\right),
\label{distr-slice}
\end{align}
where $C_2$ is a constant that can be chosen to simplify some
later equations, and $\ell$ is the ``mean free path''. For a
Gaussian, only the second cumulant is nonzero. For class BD, it
takes the form (for $N\neq 2$)
\begin{align}
\left[ \theta_{a_1 b_1} \theta_{a_2 b_2} \right]_c &= c_2
\delta_{a_1 a_2}  \delta_{b_1 b_2}, \label{2nd-cumulant}
\end{align}
where $c_2 = a/4C_2 \ell$. For a generic $K$-invariant
distribution, higher cumulants are also nonzero, and should be
expressed in terms of invariant tensors of $K$. Let us illustrate
this for class BD. In this case, for an element $k$ of $K =
\text{SO}(N)\times\text{SO}(N)$ we write
\begin{align}
k &= \begin{pmatrix} k_1 & 0 \\ 0 & k_2 \label{k1k2}\end{pmatrix},
\end{align}
where $k_i \in \text{SO}(N)$. Then as $K$ acts on $T_m$ by
conjugation, $T_j\to kT_j k^{-1}$, this gives
\begin{align}
\theta \to k_1 \theta k_2^{-1} = k_1 \theta k_2^{T},
\end{align}
and we see that the second cumulant (\ref{2nd-cumulant}) is
expressed in terms of SO($N$)-invariant Kronecker delta in the
left and right indices of $\theta$, so is invariant under
SO$(N)\times$SO$(N)$. Moreover, it is invariant under O$(N)\times
$O$(N)$. Invariant tensors under O$(N)\times$O($N$), acting in
this fashion on $\theta$, all have a similar form involving
Kronecker $\delta$s in the left and right indices. (The cumulants
\begin{align} \left[
\theta_{a_1 b_1} \theta_{a_2 b_2} \ldots \theta_{a_{2k} b_{2k}}
\right]_c \label{higher-cumulants}
\end{align}
must also be invariant under permutations of pairs $a_i$, $b_i$.)
The number of such invariant expressions increases rapidly with
the degree of the cumulant, and the $2k$th cumulant contains a
distinct parameter for each one.

However, there is also another invariant tensor of SO($N$), which
is not invariant under O$(N)$: the completely antisymmetric
Levi-Civita tensor $\varepsilon_{a_1 a_2 \ldots a_N}$. Hence,
there can be a term in the $N$th cumulant of the form
\begin{align}
c'_N \varepsilon_{a_1 a_2 \ldots a_N} \varepsilon_{b_1 b_2 \ldots
b_N}. \label{Nth-cumulant}
\end{align}
For even $N$, the $N$th cumulant will contain contributions of
both types. At higher degrees, there are also invariants that are
products of both one or more Levi-Civita tensors and Kronecker
$\delta$s. Under the additional symmetry, which corresponds to
acting on $\theta$ with $k_1$ and $k_2$, where $\det k_1 \det k_2
=-1$, the Kronecker $\delta$ function is invariant, but the
Levi-Civita tensor reverses sign. Thus the cumulant
(\ref{Nth-cumulant}) is the one of lowest degree that (if $c_N'$
is nonzero) breaks the additional symmetry, but not that under
$K$.

For class DIII, the matrix $\theta$ is imaginary and
antisymmetric. $K=$ SO$(2N)$ acts on it by conjugation.
Consequently, there are $K$-invariant cumulants containing
Kronecker $\delta$s in arbitrary pairs of indices (the lowest
being $\left[ \theta_{a_1 a_2} \theta_{a_3 a_4} \right]_c = - c_2
\bigl(\delta_{a_1 a_3}\delta_{a_2 a_4} - \delta_{a_1
a_4}\delta_{a_2 a_3} \bigr)$ with $c_2 = a/4C_2 \ell$), which are
invariant under O$(2N)$, and the lowest-order form that is
invariant under SO$(2N)$ but not under O$(2N)$ is now
\begin{align}
c'_N \varepsilon_{a_1 a_2 \ldots a_{2N}}, \label{Nth-cumulant'}
\end{align}
which is again $N$th order in $\theta$.

Thus in both classes BD and DIII, for $N>2$ the use of a Gaussian
distribution loses the symmetry-breaking aspect of the generic
distribution, and introduces an additional symmetry into the
evolution of the probability distribution with the length of the
wire. On the other hand, for $N=1$, the cumulant that breaks the
additional symmetry is degree one, and corresponds to the nonzero
mean or drift for $\theta$. For $N=2$, the symmetry-breaking term
is degree two, the same as that obtained from the invariant
Gaussian. Both terms in the cumulant can be obtained by using a
more general form of Gaussian distribution \cite{TBFM}, of the
forms
\begin{align}
P[A] \propto \exp \Biggl( -\frac{C_2\ell}{a}
    {\rm tr}\,A ^2 - \frac{C_2'\ell}{a} \!\! \sum_{a_1,a_2 \atop  b_1, b_2}
\!\!    \varepsilon_{a_1a_2}\varepsilon_{b_1b_2}
    \theta_{a_1b_1}\theta_{a_2b_2}\Biggr) \label{distr-slice'}
\end{align}
for class BD, and
\begin{align}
P[A] \propto \exp \Biggl( -\frac{C_2\ell}{a}
    {\rm tr}\,A ^2 - \frac{C_2'\ell}{a} \!\!\! \sum_{a_1,\ldots,a_4}
\!\!\!    \varepsilon_{a_1a_2a_3a_4}
    \theta_{a_1a_2}\theta_{a_3a_4}\Biggr) \label{distr-slice''}
\end{align}
for class DIII. Thus also for $N=2$, one might expect that
localization can be induced by introducing the distributions with
nonzero $C_2'$ (the corresponding effect on the DOS was observed
in Ref.~\onlinecite{TBFM}).

In the rest of this paper, we analyze explicitly the role of
symmetry breaking by the probability distribution for $T_j$,
within the FP approach, for all $N$, which means that in general
the FP equation has to be generalized to incorporate non-Gaussian
effects. We also perform an analysis (for class BD) within the
supersymmetry approach.

%%%%%%%%%%%%%%%%%%%%%%%%%%%%%%%%%%%%%%%%%%%%%%%%%%%%%%%%%%%%%

\section{Generalized Fokker-Planck equation}
\label{sec:GenFP}

In this section we first briefly review the main aspects of the
continuum limit or FP approach for the BdG symmetry classes. Then
for classes BD and DIII we argue that while, for well-behaved
distributions, the cumulants that break the additional symmetry
are lost in the usual continuum limit, in general for the analysis
of the long time behavior they should be included as higher-order
differential operators within a generalized FP equation. Then we
show that at long times, where the problem reduces effectively to
the $N=1$ case, these terms reduce to the drift term and produce
localization.

If one uses a Gaussian distribution of the form
(\ref{distr-slice}) independently for each $T_j$, then in the
continuum limit $a\to0$ one obtains a heat (diffusion) or FP
equation on $G$, which can be reduced to one on the coset space
$G/K$ (the derivation is reviewed below for a more general
distribution of $T_j$). In view of the invariance under the left
action of $K$ of both the stochastic process on $G/K$ and the
initial condition, the probability distribution on $G/K$ depends
only on the radial coordinates $x_i$, and the process can be
viewed as one on the double coset space $K\backslash G/K$ for the
joint distribution $P(x;L)$ of the $x_i$ (using the initial
condition $P(x;0) = \prod_{i} \delta(x_i)$). For the BdG classes,
this FP equation takes the form (first studied in the physical
context in Ref.~\onlinecite{BFGM}):
\begin{align}
\frac{\partial P}{\partial L} &= \frac{1}{2 \gamma \ell}
\sum_{i=1}^{N} \frac{\partial}{\partial x_i}
J \frac{\partial}{\partial x_i} J^{-1}P,
\label{eq:FPbdg}\\
J(x) &= \prod_{i=1}^{N} \sinh^{m_l} 2x_i
\prod_{i<j}^{N} \prod_{\pm} \sinh^{m_o} (x_j\pm x_i),
\label{jacobianJ}
\end{align}
where $m_l$ and $m_o$ are non-negative integers (multiplicities of
reduced roots) which depend on the symmetry of the problem ($m_o =
1$, $m_l = 0$ for class BD, and $m_o = 2$, $m_l = 0$ for class
DIII), and $\gamma = m_o(N-1) + m_l +1 $. [To obtain this form, we
have now fixed the previously-introduced parameter $C_2$ in terms
of $\gamma$, by $\gamma = 4C_2$.] The differential operator
\begin{align}
D_2 \equiv \sum_{i=1}^{N} \frac{\partial}{\partial x_i} J
\frac{\partial}{\partial x_i} J^{-1},
\end{align}
which appears in the right-hand side of Eq.~(\ref{eq:FPbdg}) is
related to the part $\Delta_2$ of the Laplace-Beltrami operator on
the space $G/K$ that is independent of the angular coordinates
(for the information on analysis on symmetric spaces, see
Refs.~\onlinecite{Helgason1}--\onlinecite{Caselle-review};
relevant facts are also given in Appendix \ref{app:diff-ops}):
\begin{align}
D_2 = J \Delta_2 J^{-1}.
\label{D2-Delta2}
\end{align}
The factor of $J$ arises because $P$ is the probability density on
$K\backslash G/K$, not on $G/K$. It is the Jacobian that appears
in the $G$-invariant integration measure over $G/K$ in terms of
the radial coordinates $x_i$ and angular coordinates in $K$. For
classes BD and DIII, $J$ and $D_2$ are invariant under
$x_1\to-x_1$.

Pertinent aspects of the solution of the Fokker-Planck equation
are reviewed in Appendix \ref{app:FP-equation-solution}. In
particular, the solution has a simple form for long wires when $L
\gg \gamma \ell$:
\begin{align}
P(x;L) &\approx \Bigl(\frac{\gamma\ell}{2\pi L} \Bigr)^{N/2}
\prod_{i=1}^{N} \exp \Bigl(-\frac{\gamma\ell}{2L}\bigl(x_i -
L/\xi_i\bigr)^2\Bigr), \label{asymptotic-solution-2}
\\
\xi_i &= \gamma \ell \bigl( m_o(i-1)+m_l \bigr)^{-1}.
\label{localization-lengths}
\end{align}
The asymptotic form (\ref{asymptotic-solution-2}) is a product of
Gaussian factors, one for each radial coordinate. This implies
that for large values of $L/\gamma\ell$, the radial coordinates
$x_i$ are with high probability well separated compared with their
fluctuations. This behavior of $x_i$ is stated precisely in the
Oseledec theorem \cite{Oseledec}, which says that the $x_i$ become
statistically-independent and Gaussian-distributed with means and
variances proportional to $L$:
\begin{align}
\la x_i \ra &= L/\xi_i, & \text{var }x_i &= L/\gamma\ell.
\label{crystallization}
\end{align}
In these relations the quantities $\xi_i$ are called localization
lengths, and their inverses are the Lyapunov exponents.

Here we note that the easiest way to understand Eqs.\
(\ref{asymptotic-solution-2}--\ref{crystallization}) is to take
the limit of large and widely separated $x_i$ directly in the
Fokker-Planck equation. As explained in Appendix
\ref{app:diff-ops}, in this limit the expression for the Jacobian
$J(x)$ simplifies, so that $
\partial \ln J/\partial x_i \to 2 \bigl( m_o(i-1)+m_l \bigr),
\label{dlnJ-asymp} $ leading to the asymptotic form of the
Fokker-Planck equation:
\begin{align}
\frac{\partial P}{\partial L} = \frac{1}{2 \gamma \ell}
\sum_{i=1}^{N} \left( \frac{\partial^2 P}{\partial x_i^2} -
2\bigl( m_o(i-1)+m_l \bigr) \frac{\partial P}{\partial x_i}
\right). \label{eq:FP-longL}
\end{align}
This is a simple diffusion equation with a constant diffusion
coefficient and a drift, and its solution (using the same initial
condition as before) is Eq.~(\ref{asymptotic-solution-2}).

The Landauer conductance (\ref{Landauer-conductance}) of a long
wire is dominated asymptotically by the smallest radial coordinate
$x_1$:
\begin{align}
{\cal G} \sim d/\cosh^2 x_1.
\end{align}
Thus, the leading behavior of both the mean and the typical values
of $\cal G$ is determined by the longest of the localization
lengths $\xi \equiv \xi_1 = \gamma\ell/m_l$, corresponding to the
smallest Lyapunov exponent. In most symmetry classes $\xi_1
<\infty$, and the leading behavior of $\cal G$ is exponentially
small, and this is referred to as strong or exponential
localization.

A distinctive feature of the FP equation for classes BD and DIII,
first investigated in the physical context in
Ref.~\onlinecite{BFGM}, is that the root systems of the
corresponding symmetric spaces do not have long roots: $m_l = 0$.
Consequently, there is no drift term for $x_1$ in
Eq.~(\ref{eq:FP-longL}), which results in $x_1$ performing a
random walk about zero. (We now understand this behavior as a
consequence of the additional symmetry of these two double coset
spaces and of the FP equations (\ref{eq:FPbdg}) under $x_1 \to
-x_1$.) Then the smallest Lyapunov exponent $1/\xi_1$ vanishes,
leading to the absence of exponential localization in long wires.
In this case, we should be a little more careful about the
treatment of $x_1$ as $L$ becomes large. We wish to assume that
all $|x_i|$ are well-separated, so that $x_2$, \ldots, $x_N$ are
large, but not necessarily that $|x_1|$ is large. For the cases of
interest, with $m_l=0$, we find that $J$ is asymptotically
independent of $x_1$, and that Eq.~(\ref{eq:FP-longL}) is still
valid in this regime (even at small $x_1$). The behavior of the
conductance $\cal G$ in these cases is characterized by a very
broad distribution, with an algebraic decay of the mean and the
variance, while the variance of $\ln {\cal G}$ is of order $L$
\cite{BFGM}:
\begin{align}
  \langle {\cal G} \rangle  &\sim d\sqrt{\frac{2\gamma\ell}{\pi L}}, &
  \text{var}\,  {\cal G} &\sim \frac{2}{3} d\langle {\cal G} \rangle,
  \nonumber \\
  \langle \ln {\cal G} \rangle &\sim -4 \sqrt{\frac{L}{2\pi\gamma\ell}}, &
  \text{var}\, \ln {\cal G} & \sim \frac{4(\pi - 2)L}{\pi\gamma\ell}.
\label{eq:localizedD}
\end{align}
Note that the mean conductance in this situation is
``super-Ohmic'', since Ohmic conductance decays as $L^{-1}$. The
wire with this behavior of the conductance was called {\it
critical} in Ref.~\onlinecite{BFGM}.

The preceding FP analysis was for the $a\to0$ limit of a Gaussian
distribution invariant not only under $K$ but under the additional
symmetry. For general $N$, especially $N>2$, we wish to include
the effect of a non-Gaussian distribution for the $T_j$ into the
analysis, since for $N>2$ these are the effects that break the
larger symmetry down to $K$ and may lead to localization. (The
following analysis also includes the cases $N=1$, $2$, in which
these effects occur already at the level of general $K$-invariant
Gaussian distributions.) If we wish to cast this into the very
convenient form of a differential equation in continuous time,
then we must first consider the $a\to0$ limit for the more general
situation of an arbitrary distribution for each time step. (This
may be a suitable place at which to mention that the use of
discrete but statistically independent increments $T_j$ can be
used to model not only processes that are actually discrete, but
also those that are continuous but not $\delta$-correlated in
time, using $a$ of order of the correlation time, provided the
correlations are not long range. In practice, the correlation time
might be of about the same order as the mean free path. As we will
see, it is in these situations that the symmetry breaking effects
for $N>2$ can show up.)

In the standard derivation of a FP equation, there is some
distribution for each (independent, identically-distributed)
increment of the random variables being considered. In the
following, we make the assumption that this distribution has
finite moments of all degrees. The change in the probability
density for the variables at time $m+1$ from that at time $m$ at
any point is given by the master equation, and contains two terms,
one for reduction of the probability by transitions to all other
values, the other for the increase by transitions from other
values (we give only a sketch of this derivation, as it can be
found in many textbooks, and adapted to the present case without
difficulty). Since the distribution of increments is narrow, the
increments in the variables are typically small, and in the second
term the probability distribution for time $m$ can be Taylor
expanded in the increments, about the values at time $m+1$. The
convolution integral over these increments then reduces in each
term to a multiple partial derivative of the product of a moment
of the distribution for the increment and the probability
distribution. The zeroth-order such term cancels the first term in
the master equation. The result, in which we have not yet taken
any limit $a \to 0$, is a generalization of the FP equation which
is a first-order finite difference equation in the time variable
$m$, and all orders in partial derivatives in the variables. Such
an expansion for a finite increment in the variables in terms of
derivatives of the distribution is known as a Kramers-Moyal
expansion, see, for example, Ref.~\onlinecite{Risken}.

In the present case, using a $K$-invariant distribution of the
increments (described by the matrix $A$), each term is a
differential operator on $G/K$ that is invariant under the left
action of the whole of $G$. Each operator of order $k$ has a
coefficient that is related to a $k$th moment of the distribution
of increments. By approximating the time step of length $a$ by a
first derivative, we obtain a generalized form of FP equation for
the probability distribution $P(\theta;L)$ on $G/K$, where the
variables $\theta$ introduced earlier also serve as the
coordinates on $G/K$, with $\theta=0$ being the origin, which
corresponds to the transfer matrix $M$ being the identity in $G$.
The $G$-invariant differential operators on symmetric spaces have
been extensively investigated, see for example
Ref.~\onlinecite{Helgason2}. Because sums, differences, and
products of invariant operators are also invariant, the
$G$-invariant operators form an algebra, which turns out to be
commutative, and in fact all the operators can be obtained as
linear combinations of products of a finite set of operators. For
both classes BD and DIII, these operators can be labelled
$\Delta_2$, $\Delta_4$, \ldots, $\Delta_{2N-2}$, each of which has
order given by its subscript, and $\Delta_{p_N}'$, which is of
order $N$. Thus the
generalized FP equation can be written as %
\begin{align} %
a\frac{\partial P(\theta;L)}{\partial L} = \!\!\!
\sum_{n_1,\ldots,n_N} \!\!\!  \frac{m_{n_1,\ldots,n_N}}{k_{\rm
tot}!} \Delta_{p_N}'^{n_N} \biggl(
\prod_{k=1}^{N-1}\Delta_{2k}^{n_k} \biggr) P(\theta;L),
\label{generalized-FP}%
\end{align} %
where the coefficients $m_{n_1,\ldots,n_N}$ are essentially
moments of degree $k_{\rm tot} = \sum_{k=1}^{N-1} 2k n_k + N n_N$
of the increments $\theta$, except for $m_{0,\ldots,0}$ which is
zero due to probability conservation. [Strictly, these are the
moments of $\theta$ minus its mean, except for the coefficient of
the first-order operator, if any, which is the mean of $\theta$.
But the mean is zero except in the $N=1$ case, which as we saw can
be handled directly.] The sum is over nonnegative $n_k$.

The form of the differential operators is easy to understand if we
examine the region near the origin. For class BD, invariant
differential operators can be constructed from $\partial/\partial
\theta_{ab}$. Invariance under $K=$ SO$(N)\times$ SO$(N)$ implies
that near $\theta=0$, the above operators reduce to %
\begin{align} %
\Delta_{2k} &= \!\!\! \sum_{a_1,\ldots,a_{k} \atop
b_1,\ldots,b_{k}} \!\!\! \frac{\partial}{\partial \theta_{a_1b_1}}
\frac{\partial}{\partial \theta_{a_1b_2}}\frac{\partial}{\partial
\theta_{a_2b_2}}\cdots
\frac{\partial}{\partial \theta_{a_k b_1}}, \label{localDelta-k}\\
\Delta_{p_N}'&= \!\!\! \sum_{a_1,\ldots,a_{N} \atop
b_1,\ldots,b_{N}} \!\!\! \varepsilon_{a_1\ldots
a_N}\varepsilon_{b_1\ldots b_N}\frac{\partial}{\partial
\theta_{a_1b_1}}\cdots \frac{\partial}{\partial
\theta_{a_Nb_N}}.\label{localDelta-p}%
\end{align} %
Here we see that the first $N-1$ operators resemble traces, and
were constructed using Kronecker $\delta$s, while the last used
the Levi-Civita tensor $\varepsilon$. Hence, the first $N-1$ are
invariant under the larger symmetry [the action of any element in
O$(N)\times$ O$(N)$], while $\Delta_{p_N}'$ is odd under elements
of O$(N)\times$ O$(N)$ that have determinant $-1$. For class DIII,
one has, more simply,
\begin{align} %
\Delta_{2k} &= \frac{1}{2^{2k}} \, {\rm tr}
\left(\frac{\partial}{\partial
\theta}\right)^{2k} , \\
\Delta_{p_N}' &= \frac{i^N}{2^N} \!\!\!\!\!
\sum_{a_1,\ldots,a_{2N}} \!\!\! \varepsilon_{a_1\ldots
a_{2N}}\frac{\partial}{\partial
\theta_{a_1a_2}}\cdots \frac{\partial}{\partial \theta_{a_{2N-1}a_{2N}}},%
\end{align}
with similar transformation properties, this time under O$(2N)$.
We emphasize that in both cases these forms are only valid for
small $\theta$, and that for general positions on $G/K$ the
expressions become more complicated, in such a way that the
operators are invariant under $G$ (although, in local coordinates
around any point, they would take the same above form, due to the
symmetry of the symmetric space).

The operator $\Delta_2$ is the Laplace-Beltrami operator on $G/K$
in both cases. For the special case $N=2$, $\Delta_{p_N}'$ is also
second order. It is a theorem that for an irreducible symmetric
space $G/K$, the Laplace-Beltrami operator is the unique
second-order invariant differential operator, up to multiplication
by a constant. For $N=2$, the symmetric spaces for transfer
matrices in classes BD and DIII cease to be irreducible. This is
basically because of the isomorphisms of Lie algebras so(4)
$\cong$ sl(2)$\times$sl(2), and the symmetric spaces likewise
become direct products \cite{Helgason1}. There is then a distinct
Laplace-Beltrami operator for each factor in this product (each
operator involving only derivatives along the directions lying
within the factor space). The same essential point was noticed
earlier in Ref.~\onlinecite{TBFM}. [Incidentally, the transfer
matrix spaces for all three chiral symmetry classes for $N>1$ are
also reducible, and contain the real line as a direct factor (the
position on this line is described by $\sum_i x_i$). Consequently,
for all three cases, there are two invariant second-order
operators (as pointed out for the orthogonal and unitary cases in
Ref.~\onlinecite{BMF}), and a drift term along the real line
(i.e.\ a first-order operator) is also invariant under the simply
connected group $G$, but breaks the discrete symmetry which sends
$\theta\to-\theta$, or in terms of the radial variables,
$x_i\to-x_i$ for all $i$. In a chain with off-diagonal randomness
in the nearest-neighbor hopping, the drift term is produced by
staggering the magnitude of the mean hopping
\cite{balents-fisher,BMSA}.]

Due to the invariance of the evolution and initial conditions
under the subgroup $K$, only the radial parts of the operators
$\Delta_{2k}$ and $\Delta_{p_N}'$ will be needed in the following.
In terms of the radial parts as defined in Appendix
\ref{app:diff-ops}, $\Delta_{p_N}' = N!\Delta_{p_N}$. The angular
dependence of $P(\theta;L)$ is trivial, and we can reduce the
equation to one for the probability density in only the $x_i$
variables, $P(x;L)$. Because $P$ is a probability density, the
relation is $P(x;L)=JP(\theta;L)$. The generalized FP equation for
$P(x;L)$ takes the same form as Eq.~(\ref{generalized-FP}), but
with $D_{2k}$ and $N! D_{p_N}$ in place of $\Delta_{2k}$ and
$\Delta_{p_N}'$, where %
\begin{align}%
D_{2k} &= J\Delta_{2k}J^{-1},\nonumber\\
D_{p_N} &= J\Delta_{p_N} J^{-1}.\label{conjugation}
\end{align}

We can now describe the fate of the terms of more than second
order in the generalized FP equation, as $a\to 0$. For any
probability distribution of $T_j$ which has finite second cumulant
with coefficient $c_2$ for $\theta$, we define the mean free path
$\ell$ through Eq.~(\ref{2nd-cumulant}), where $C_2$ depends on
the symmetry class, and is chosen so that the coefficient of the
second order operator in the FP equation takes the standard form,
Eq.~(\ref{eq:FPbdg}). We are only interested in $a < \ell$. If
$\theta$ has nonzero mean, which can occur only for $N=1$, then we
take this mean to be a number times $a/(\gamma\ell)$. The moments
$m_{n_1,\ldots,n_N}$ (other than the mean) with total degree
$k_{\rm tot} =\sum_{k=1}^{N-1} 2k n_k + N n_N$ are then of order
$(a/\gamma\ell)^{k_{\rm tot}/2}$. If we take the limit $a \to 0$
with $\gamma\ell$ fixed, then all operators of higher order than
second drop out, and the usual FP equation is obtained. For $N>2$,
all terms that break the additional symmetry are lost in this
limit. For $N=1$ and $2$, a drift term, and an additional
second-order term respectively survive in the limit, and these do
break the symmetry unless their coefficients are tuned to zero. It
has been noticed previously that localization ensues when the
drift term is present for $N=1$, as discussed above \cite{Chalk},
and that the non-critical form of the DOS,
Eq.~(\ref{DOS-varying}), results when the additional second-order
term is included \cite{TBFM} for $N=2$ . [Similarly, for the three
chiral classes, the $N$ transitions, at each of which the behavior
is critical, can be reached within the FP approach by tuning the
coefficient of the drift term\cite{BMSA}.]

It is instructive to realize that the vanishing of all
higher-than-second-order terms in the limit $a\to0$ is connected
with the central limit theorem, and also to interpret it in terms
of the renormalization group (RG). When the typical values of
$\theta_{ab}$ for each time step are small (compared with
$1/\sqrt{N}$), then in the product of $T_j$ over any time interval
that is not too long (less than about $\ell$ --- not $\gamma\ell$,
as we explain below), to a good approximation the $\theta$s simply
add (as matrices). $G/K$ is a smooth manifold, and has a finite
radius of curvature at any point. Over this time interval, the
random position on $G/K$ has been displaced by an amount less than
the radius of curvature. Hence we are essentially looking at a
random walk in a flat space. The central limit theorem for sums of
independent, identically-distributed random variables then tells
us that the sum becomes Gaussian distributed in the limit of many
variables, provided the mean and variance for each variable are
finite. This means that the probability density for the sum
effectively obeys a FP equation with no higher order terms over a
long enough time interval. In this flat space case, this can be
understood in terms of the generalized FP equation as saying that
as the time gets longer, the distribution spreads, and the values
of the higher derivatives become smaller relative to the second
derivative, and so can be neglected. In renormalization group
language, the terms of higher order than second are irrelevant (in
this flat space regime), as they become less important at larger
time scales. However, if we reverse this process and go to shorter
time scales, the higher derivatives get larger. If we take the
full generalized FP equation literally and use it for the
evolution of an initial $\delta$ function for arbitrarily short
times, then it ceases to make sense. Of course, the derivation
from discrete time steps implies that the equation for time step
$a$ certainly cannot be used for times less than $a$. It
effectively describes the process ``at the time scale $a$''. The
fact that the irrelevant operators have coefficients that contain
positive powers of $a$ (the short-time or ultraviolet cutoff) is
typical for irrelevant operators. As always in the RG,
``relevant'' and ``irrelevant'' refer to linearized scaling
behavior near a particular fixed point. In the present case, the
fixed point is the ballistic fixed-point, which is described by
the FP equation, or by using a Gaussian distribution for the
increments and taking the $a\to0$ limit. The $a\to0$ limit can be
taken only if we are prepared to drop the effects of the
irrelevant operators.

We now examine the behavior at longer times. As its name implies,
the ballistic fixed-point describes very short times where there
is no scattering. The mean free path $\ell$ is the time scale at
which scattering, which corresponds to curvature of $G/K$, first
shows up. We are interested in $a$ small, $a\ll \ell$, and all the
coefficients of the irrelevant operators can be thought of as
small, though we do not need to set them to zero. Over these time
intervals, the central limit theorem applies, and the FP equation
gives a reasonable description of the evolution of $P(x;L)$. For
$L\ll \ell$, all $x_i$ are small, and ${\cal G}\simeq dN$. We
should note that the squares of the $x_i$s are the eigenvalues of
$\theta\theta^T$, and so have typical values of order $N$ times
those of the squares of elements $\theta_{ab}$. As the time $L$
passes through the localization lengths $\xi_N < \xi_{N-1} <
\cdots < \xi_2$, typical values of the corresponding $x_i$ pass
through the vicinity of 1 and begin to grow linearly (the smallest
such length $\xi_N$ is close to $\ell$). Their contribution to
$\cal G$ therefore drops, and we are in the Ohmic regime in which
the mean and typical $\cal G$ are of order $dN\ell/L$ ($N\ell/L$
is the number of channels not yet localized at time $L$). Here
there are also fluctuations in $\cal G$. This crossover behavior
is ``universal'', even at finite $N$, in the sense that the
irrelevant operators can be dropped; the universality arises, as
usual in the RG, from the RG flow passing close to a fixed-point,
in this case the ballistic fixed point which is reached as $L\to0$
provided the irrelevant operators are dropped. However, this
universality may not be of great interest physically. If one also
takes the limit $N \to \infty$ with $\gamma\ell$ fixed, so $\ell
\to 0$, before taking $L \to 0$, then one arrives at a
``diffusive'' fixed point. That is, if $N$ is large but finite,
then the RG flow from the ballistic fixed point approaches close
to the diffusive fixed point. This universality in the Ohmic
regime when the diffusive fixed point is approached as $L\to0$ is
what is usually meant when the term ``universal'' is used in
mesoscopic physics for the behavior of the conductance in a 1D
system. The diffusive fixed point will not concern us here, but we
point out that the higher-order differential operators are also
irrelevant at this fixed point, and so are dropped to obtain this
universal behavior also. Then not only the Ohmic regime, but the
whole crossover that begins from the diffusive fixed point at
$L\to0$ is universal, and can be studied as in
Ref.~\onlinecite{BFGM}.

When the time $L$ passes approximately the inverse of the last
nonzero Lyapunov exponent, that is $L\simeq \xi_2\simeq
\gamma\ell/m_o$, then only $x_1$ is still not growing linearly. If
one uses the FP Eq.~(\ref{eq:FPbdg}), then this is where the
asymptotic behavior termed critical sets in. Thus the Ohmic regime
is a crossover from the diffusive fixed point to a critical fixed
point, and $\xi_2$ is our estimate for the crossover time,
$\xi_\times\sim \xi_2 \sim \gamma\ell$. In most other symmetry
classes, localization would set in at times larger than $\xi_1$,
so that the localization length $\xi$ is about $\gamma\ell$, in
agreement with RG analysis in 1D. In the present case, the
analysis of the RG is similar, but the crossover is to a critical
rather than a localized fixed point.

We are concerned with the effect of the symmetry-breaking
operators in the generalized FP equation on the critical fixed
point. We assume that the coefficient of the leading such term
($D_{p_N}$) is small, but not zero. Then it, and all other terms
of orders higher than second, can be dropped when studying the
crossovers from ballistic to diffusive to critical. The next
question is whether they are important (relevant) at the critical
fixed point.

The main points can be understood by retaining in the right-hand
side of the generalized FP only the terms containing the
differential operators $D_2$ and $D_{p_N}$, so that
\begin{align}
\frac{\partial P(x;L)}{\partial L} &= \frac{1}{2\gamma\ell} D_2 P
+ (-1)^N\frac{c'_N}{a} D_{p_N} P, \label{generalized-FP'}
\end{align}
(note that $m_{0,0,\ldots,1}=(-1)^N c'_N$). Because we are in the
regime where all $x_i$ except $x_1$ are large, and because the
asymptotic behavior for $c'_N=0$ is dominated by $x_1$, we will
attempt to ``integrate out'' these $x_i$. In Appendix
\ref{app:diff-ops}, we determine explicitly the asymptotic forms
${\widetilde D}_{2k}$ and ${\widetilde D}_{p_N}$ of the operators
$D_{2k}$ and $D_{p_N}$ for classes BD and DIII, in the regime
$|x_1|\ll x_2 \ll\cdots\ll x_N$ (but no other assumptions on
$x_1$). Then we define the reduced probability density
$P_1(x_1;L)$ for $x_1$ by integrating $x_i$, $i\geqslant 2$ over
the range $|x_1| \leqslant x_2\leqslant \cdots\leqslant x_N$. An
equation for $P_1$ can be obtained by integrating the generalized
FP equation over the same range. By using the asymptotic form of
$P(x;L)$, Eq.~(\ref{asymptotic-solution-2}), (in the derivation of
which, the effects of $c'_N$ and other higher-order operators are
negligible), we find that in the $\widetilde D$'s, the
$\partial/\partial x_i$s for $i\geqslant 2$ can be replaced by
zero to obtain the corresponding differential operators in $x_1$
only, which act on $P_1$. In particular, ${\widetilde D}_{p_N}$
becomes $\propto\partial/\partial x_1$, and still breaks the
symmetry $x_1\to -x_1$. The resulting equation (for either class
BD or DIII) is
\begin{align}
\frac{\partial P_1}{\partial L} &= \frac{1}{2 \xi_\times}
\frac{\partial^2 P_1}{\partial x_1^2} + \kappa \frac{\partial
P_1}{\partial x_1}. \label{generalized-eff-theory}
\end{align}
Here we used
\begin{align}
\xi_\times &= \gamma \ell,  \label{xitimes}\\
\kappa &= - m_o^{N-1}(N-1)!\frac{c'_N}{a}. \label{kappa}
\end{align}
Equation (\ref{generalized-eff-theory}) has no explicit
$N$-dependence, and in particular is precisely the form obtained
for $N = 1$, on dropping operators higher than second order. Also,
this justifies the use of the term ``critical fixed point'' when
$\kappa = 0$, since the equation allows a scaling analysis, with
$x_1\sim (L/\xi_\times)^{1/2}$. This fixed point reproduces the
critical asymptotic behavior of $\cal G$ or $\ln \cal G$, of
course.

Hence, the symmetry-breaking terms produced by disorder that is
not invariant under the additional symmetry are relevant at the
critical fixed point, and there is a phase transition at
$\kappa=0$ between two phases. In the RG, such operators that are
irrelevant at a fixed point in the ultraviolet (such as the
ballistic or diffusive fixed points), but relevant at a fixed
point at a larger time scale (in the infrared) are called {\it
dangerously irrelevant}, because even though they are irrelevant
at the ultraviolet fixed point, they cannot be set to zero without
some important physics being lost. For $\kappa \neq 0$,
Eq.~(\ref{generalized-eff-theory}) shows that the Lyapunov
exponent for $x_1$ is $|\kappa|$, so that
\begin{align}
\xi = \frac{1}{|\kappa|}\label{kappa'}
\end{align}
describes the localization length when the typical (i.e.
high-probability) behavior of $ -\ln \cal G$ is growing linearly
as $2L/\xi$. The behavior at the transition is universal and
described by Eq.~(\ref{generalized-eff-theory}) provided $L$ and
$\xi\gg \xi_\times$, {\em even if $N$ is finite or small so that
the crossover from the diffusive to the critical fixed point is
not fully universal}. In this regime, the probability density of
$y\equiv(-\frac{1}{2}\ln {\cal G})/(L/\xi_\times)^{1/2}\geqslant
0$ approaches a universal function of $y$ and $\xi_\times
L/\xi^2$, because $\ln {\cal G}-\ln(4d)$ is approximately
$-2|x_1|$, except near ${\cal G}\simeq d$, but that region has
negligible probability for large $L$. The
precise result is%
\begin{align} %
P(y) &= \frac{1}{\sqrt{2\pi}}\exp\left\{-\frac{1}{2}\left[y -
\left(\frac{\xi_\times L}{\xi^2}\right)^{1/2}\right]^2\right\}
\nonumber\\
& \quad + \frac{1}{\sqrt{2\pi}}\exp\left\{-\frac{1}{2}\left[y +
\left(\frac{\xi_\times L}{\xi^2}\right)^{1/2}\right]^2\right\}.
\label{scalG}
\end{align}
The regime of linear growth of the typical value of $-\ln \cal G$
sets in at the time scale $L\sim \xi^2/\xi_\times$, which is much
larger than $\xi$ under our assumptions that define the scaling
regime. At smaller time scales, it grows only as
$C\sqrt{L/\xi_\times}$, for some constant $C$, as described for
the critical $\kappa = 0$ case using the $N$ channel FP equation.

All the moments of $\cal G$ are of order $e^{-(\xi_\times
L)/(2\xi^2)}\sqrt{\xi_\times/L}$, throughout the critical regime.
(This behavior differs from the ``log-normal'' behavior in a
one-channel wire in most other symmetry classes, because usually
the analogs of $\xi_\times$ and $\xi$ are proportional.) Thus,
while $\xi_{\rm typ}=\xi=1/|\kappa|$ describes the rate of
linear decay of the typical log conductance,%
\beq %
\xi_{\rm mean} = 4\xi^2/\xi_\times \label{xi-mean}%
\eeq%
describes the exponential decay of the mean conductance. Similar
behavior $-L/\xi_{\rm typ}$ and $e^{-L/\xi_{\rm mean}}$ can be
inferred for the typical behavior of the logarithm, and for the
mean behavior, respectively, of the eigenfunctions at zero energy,
as these determine the transmission through the system. This
behavior also applies to mean values of the Green's functions, as
we will see in the following section. The mean localization length
is much longer then the typical one, because the average
transmission is dominated by the rare events that produce
$|x_1|<1$.

We can check the effect of all other terms in the generalized FP
equation in the same way. We find that in asymptopia, all
${\widetilde D}_{2k}$ reduce to operators $
\partial^{2k}/\partial x_1^{2k}$. At the critical fixed point,
operators of order higher than two are again irrelevant, and the
only other term that should be kept is a correction of second
order coming from $({\widetilde D}_{p_N})^2$, which gives a small
correction to $1/\xi_\times$. As an aside we mention that a
similar analysis could be done in the (Ohmic) crossover regime, by
integrating out only the subset of $x_i$ that typically are
growing exponentially (this means introducing a somewhat arbitrary
cutoff on $i$, since there is not a clear separation of large and
small $x_i$ in this case, unlike in the vicinity of the critical
fixed point). This thinning out of degrees of freedom, a typical
method in the RG, would show explicitly how the RG acts in this
regime. The variables $x_i$ that are growing linearly correspond
to channels in which transmission is decaying exponentially, like
massive degrees of freedom in a field theory.

We can combine the forms obtained here for the mean and typical
localization lengths with the values of the moments that would be
obtained for the non-Gaussian probability distribution of each
$\theta_j$ at the lattice scale $a$. If the distribution has a
fixed form that scales with the width $\sim (a/\gamma
\ell)^{1/2}$, with a parameter $h$ for the strength of the leading
symmetry-breaking cumulant in these units ($h$ stays fixed as
$a\to0$), so that $c'_N = h(a/\gamma\ell)^{N/2}$ for $N \geqslant
2$, then we find that
\begin{align}
\frac{\xi_{\rm typ}}{\xi_\times} &= \frac{1}{m_o^{N-1}(N-1)!|h|}
\Bigl(\frac{a}{\gamma\ell}\Bigr)^{-(N/2-1)}, \label{xi-typical}
\end{align}
while $\xi_{\rm mean}/\xi_\times = 4(\xi_{\rm typ}/\xi_\times)^2$.
This illustrates how the (dangerous) irrelevance of the
symmetry-breaking perturbation at the ballistic and diffusive
fixed points may translate into very large localization lengths,
especially as $a/\gamma\ell \to 0$ with $N$ large but fixed, and
$\xi_\times$ and $h$ fixed. Note that the limit $N\to\infty$ with
$\xi_\times$ fixed is appropriate for studying the diffusive
regime and the crossover; the localization lengths appear to
vanish in this limit if $h$ and $a/\gamma\ell$ are held fixed,
however the analysis assumed that $\xi_{\rm typ}\gg \xi_\times$
(i.e.\ small $|h|$). That is, if $|h|$ is sufficiently large,
localization will occur without the RG flow approaching the
vicinity of the critical fixed point analyzed above.

%%%%%%%%%%%%%%%%%%%%%%%%%%%%%%%%%%%%%%%%%%%%%%%%%%%%%%%%%%%%%%%%%

\section{Supersymmetry solution}
\label{sec:SUSY}

In this section, we discuss the application of a supersymmetry
method to the problem of localization in a superconducting wire in
class BD. The details of the calculations are provided in Appendix
\ref{app:SUSY}, and here we only give a summary of the results.
The technique is effectively a finite-$N$ analog of the 1D
supersymmetric nonlinear $\sigma$ model, which arises in the
$N\to\infty$ limit. The forward and backward directions along the
wire are referred to here as ``up'' and ``down'', and after
averaging over disorder in the subgroup $K$, the states are in a
tensor product space of two ``superspins'', which are the
representation spaces of the superalgebra osp($2|2$), $R$ and
$\Rb$ for up and down, respectively. After averaging, these are
coupled by the backward scattering ($\theta$) part of the transfer
matrix $M$. The usual Gaussian disorder, invariant under the
additional symmetry, produces a coupling of the superspins that
corresponds to the usual Heisenberg coupling familiar from
magnetism. The leading symmetry-breaking cumulant produces an
additional term with the coefficient $\kappa$, given by Eq.\
(\ref{kappa}). When it is nonzero, it produces localization. A
dimensionless parameter $|\lambda| = \xi_\times/(2\xi_\text{typ})$
that enters the calculations in Appendix \ref{app:SUSY} is
proportional to $|\kappa|$.

The properties of the effective superspin problem are easily
translated to the properties of the disordered wire. In
particular, the mean localization length $\xi_{\rm mean}$ in the
wire is inversely proportional to the energy gap above the ground
state at zero ``energy'' (conjugate to ``time'', i.e. distance
along the wire) in the superspin problem. The DOS in the wire is
proportional to the expectation value (in the ground state) of the
``staggered magnetization'' for the two superspins, as a function
of true energy. We have solved the superspin problem and have
found its eigenstates and the spectrum. (We present results for
$N$ even only. The case $N=1$ was done previously \cite{grl1},
using results from Ref.~\onlinecite{balents-fisher}.) The excited
states are separated from the ground state by a gap that turns out
to be
\begin{align}
E_{\text{gap}} = 2/\xi_\text{mean}, \label{energy-gap}
\end{align}
where $\xi_\text{mean}$ is again given by Eqs.\ (\ref{xi-mean},
\ref{xitimes}--\ref{kappa'}) for $m_o = 1$. This implies that all
moments of $\cal G$ will again decay as $\sim e^{-2L/\xi_{\rm
mean}}$ (when $E_{\text{gap}}$ is nonzero), confirming the result
of the previous section by a very different method.

We also determine the low-energy DOS $\nu(\epsilon)$ in the wire
both at and near the critical point. At criticality ($\lambda =
0$), we obtain Dyson's singularity (\ref{DOS-Dyson}). Close to the
critical point ($|\lambda| \ll 1$) we get the divergent DOS as in
Eq.~(\ref{DOS-varying}) with exponent $\delta = 1 - 4|\lambda|$.
At criticality but at finite $\epsilon$, the behavior can be
characterized in terms of a typical localization length $\xi_{\rm
typ}\sim |\ln\epsilon|$ as $\epsilon\to0^+$, and a mean
localization length $\xi_{\rm mean}\sim \ln^2|\epsilon|$. This
distinction has been emphasized by D. Fisher in his study of
random spin chains \cite{dfisher}. The same forms that we find
here were obtained using supersymmetry techniques in
Ref.~\onlinecite{balents-fisher} for a one-channel model in the
chiral orthogonal class (BDI in AZ classification). We comment
further on the relation of the transitions in 1D in different
symmetry classes in Section \ref{sec:superu} below. It is
interesting that two analogous length scales emerged rather easily
from the transfer matrix analysis (especially if one considers
only $N=1$). For the random hopping problem (class BDI with one
channel), an approach using the transfer matrix and a scaling
argument \cite{ziman} reproduced Dyson's DOS at criticality. It
appears that one can think of the DOS integrated from $0$ to
$\epsilon$ (which can be obtained from the transfer matrix
\cite{thou}) as being of order $1/\xi_{\rm mean}$ by scaling, as
it is really a mean Green's function property; then
differentiating reproduces Dyson's result.

It was emphasized by Bocquet {\it et al.} \cite{BSZ} that the
target space of the nonlinear $\sigma$ model for classes BD and
DIII is disconnected (it is OSp$(2|2)/$U$(1|1)$ for the simplest
case in class BD, and the bosonic sector is O(2)/U(1), which has
two components). Thus there can be domain walls at which the field
jumps from one component to the other. In terms of superspins, the
representation $R$ splits into subspaces called $R_0$ and $R_N$ in
Appendix \ref{app:SUSY}, which are irreducible representations of
the superalgebra osp$(2|2)$ (there is a similar decomposition of
$\Rb$). Under the action of the superalgebra, the orbits of the
highest or lowest weights in these subspaces form supermanifolds
which are the two connected components of OSp$(2|2)/$U$(1|1)$, so
that states in these subspaces correspond to being on one or other
component of the supermanifold. The interesting point is that the
term that breaks the additional symmetry produces a term in the
superspin Hamiltonian that flips between these two subspaces, and
thus acts as a domain wall as a function of time (the domain wall
is simply a point in 1D). Hence we can conclude that domain walls
are essential to produce localization. This has been proposed
before \cite{BSZ,Chalk}, but now we have explicit results for
arbitrary numbers of channels in 1D, and in our framework we know
the precise form of the effect the domain wall produces (though we
have not analyzed the $N\to\infty$ limit).

%%%%%%%%%%%%%%%%%%%%%%%%%%%%%%%%%%%%%%%%%%%%%%%%%%%%%5

\section{Random fluxes}
\label{sec:flux}

So far, our analysis has been performed in the continuous model
introduced in Section \ref{sec:symmetry}, in which the transfer
matrices $T_j$ that are multiplied together for each time step are
in the connected group $G$, and the disorder is weak so each $T_j$
is close to the identity. But the symmetry conditions that define
classes BD and DIII do not impose this continuity. They allow the
transfer matrix, or its increments $T_j$, to belong to a larger
group $\widehat{G}$ in each case, with $\widehat{G}=$ O$(N,N)$ for
class BD, $\widehat{G}=$ O$(2N,\mathbb{C})$ for class DIII. These
groups include matrices of determinant $\pm 1$, and so are not
connected. Such transfer matrices may then be expected to arise
when the model is not continuous, for example lattice (or
tight-binding) models. The negative determinant may be interpreted
as saying that when a particle propagates around a closed loop
(going first forwards along the wire, then backwards by a
different path), it may encircle a magnetic flux of half a
quantum, often referred to as a flux of $\pi$. Thus we term this
the inclusion of random fluxes (of $\pi$) in the model (it would
perhaps be more accurate to call these random vector potentials).
In this section, we analyze the effects of these. The analysis is
straightforward in view of the previous results.

It is convenient to consider the following simple model, that is
based on the continuous one used up to now. We use a small density
(along the wire) of random fluxes. That is, generically the
evolution is by $T_j$s in $G$ with weak disorder, distributed as
before. But at random times, we also insert a matrix of negative
determinant. Such a matrix could be an arbitrary element of
$\widehat{G}$. One may wish to view it as a $T_j$ close to the
identity, times some element with a negative determinant. As we
factor out the $T_j$ close to the identity anyway, we may as well
choose the extra matrix to belong to a maximal unitary subgroup
$\widehat{K}$ of $\widehat{G}$, which contains $K$. Relative to
the same basis we have used throughout, which refers to the
decomposition of $M'$ into in and out subspaces, the natural
choice for $\widehat{K}$ is O$(N)\times$O$(N)$ or O$(2N)$ for
classes BD and DIII, respectively (other choices for $K$ or
$\widehat{K}$ are isomorphic to these). With this choice, our
remarks above about picking up a phase $-1$ on propagating around
some loops describe exactly what takes place, up to the effect of
multiplication by some element of $K$ (if we neglect the disorder
outside $K$ which is small over a small time-step).

We note that for class BD, $\widehat{K}$ has four connected
components, as may be seen by writing its elements in the form of
Eq.~(\ref{k1k2}) but with $k_1$ and $k_2$ in O$(N)$. Then the four
choices of signs for the determinants of $k_1$ and $k_2$ label the
four components. For class DIII, there are just two connected
components of $\widehat{K}$. Because O$(N)$ is a semidirect, not a
direct, product of SO$(N)$ and ${\mathbb Z}_2$ when $N$ is even,
there is no unique (i.e.\ invariant under $K$) choice of negative
determinant matrix in $\widehat{K}$ for class DIII or for class BD
for $N$ even. For class BD with $N$ odd, there are two possible
natural choices. If we write elements of $\widehat{K}$ in the form
of Eq.~(\ref{k1k2}) but with $k_1$ and $k_2$ in O$(N)$, then we
can take one of $k_1$, $k_2$ equal to $I_N$, the other equal to
$-I_N$.

As time evolves (i.e.\ $L$ increases), the transfer matrix $M$
evolves on $\widehat{G}$ by diffusion as before, but now also with
occasional jumps between the different connected components of
$\widehat{G}$, due to the occasional insertion of a negative
determinant element of $\widehat{K}$ into the product of transfer
matrices for small steps. Modulo multiplication on the right by
elements of $K$, $M$ evolves on the coset space $\widehat{G}/K$,
which has four or two connected components for classes BD and DIII
respectively, usually by diffusion but with occasional jumps to
another component. Modulo the left action of $K$ on these
symmetric spaces, evolution is on four (two) copies of the double
coset space $K\backslash G/K$ for class BD (DIII, respectively).
For class BD, the distinction between components with the same
sign for the determinant appears to be unimportant, so we
disregard it. The important point is that there is a
correspondence between points on the two copies of $K\backslash
G/K$, and when a jump occurs, it is between two corresponding
points. If we parametrize the two connected spaces under this
correspondence by the usual coordinates $x_i$, then the only
difference between evolution on the two components is that in the
generalized FP, which applies between jumps, the terms that break
the symmetry $x_1\to -x_1$ have opposite signs on the two
components.

We may now consider the effect of the random fluxes on the wire
over very long time (length) intervals that contain many jumps. If
the jumps are correlated in pairs, so that the system spends most
of its time on one component (say, the one in which $M$ is in the
identity component), then there will be a net effect of the
symmetry breaking terms. In the critical regime, there will be a
net drift term. In this case, a transition as we have described it
in the previous sections will still occur, and we expect the same
universal properties. This is what occurs in the lattice models
considered in Refs.~\onlinecite{Motrunich,merz}.

On the other hand, if the jumps are uncorrelated, or more
generally if equal time is spent on both components, then the
average drift will be zero. In this case, localization is
suppressed and the system remains critical. (We note that both
types of behavior were observed in 1D $N=1$ models in
Ref.~\onlinecite{Chalk}.) In this case, the results can also be
obtained by another approach. After moderate time intervals, the
transfer matrices $T_j$ become uniformly distributed over the {\em
disconnected} angular group $\widehat{K}$. We can treat this
situation as we did for the connected group $K$ in the continuous
model, which leads us to consider the coset spaces
$\widehat{G}/\widehat{K}$ and the double cosets
$\widehat{K}\backslash \widehat{G}/\widehat{K}$. These are
connected spaces, and $\widehat{G}/\widehat{K}$ is the same as
$G/K$. The Weyl group inside $\widehat{K}$ now acts on the radial
coordinates $x_i$ by permutations and arbitrary sign changes, so
they lie in the Weyl chamber $0<x_1<x_2<\cdots<x_N$, which is the
quotient of the previous $C$ by the operation $x_1\to -x_1$; this
parametrizes $\widehat{K}\backslash \widehat{G}/\widehat{K}$. In
this case, the generalized FP equation on
$\widehat{G}/\widehat{K}$ can contain only differential operators
invariant under $\widehat{K}$, and so $\Delta_{p_N}$ is not
allowed, and its place among the generators of the algebra of
invariant differential operators is taken by $\Delta_{2N}$ which
is of order $2N$. It is now clear from our preceding analysis that
in this case there can be no relevant perturbation of the critical
fixed point, and the system remains critical, with behavior as
described in Ref.~\onlinecite{BFGM}.

We can also see how these effects would emerge in the
supersymmetry framework for class BD (see Appendix \ref{app:SUSY}
for terminology). If the angular variables must be averaged over
$\widehat{K}$, then this produces a constraint that the states on
the up (down) sites must be in $R_0$ ($\Rb_0$). Hence there can be
no $\lambda H_1$ term in the Hamiltonian, and in this case all
states will be extended. Thus, suppressing domain walls suppresses
localization. (Again, this was discussed previously for 2D, and
for $N=1$ in 1D\cite{BSZ,Chalk}.)

For the chiral symmetry classes, only the chiral orthogonal class,
BDI, possesses a disconnected transfer matrix group $\widehat{G}=$
GL$(N,{\mathbb R})$, while $\widehat{K}=$ O$(N)$, and both of
these groups have two connected components. However, unlike the
cases of classes BD and DIII, in this case $\widehat{K}$ acting by
conjugation on the $T_j$ does not reverse the sign of all $x_i$
(the desired additional symmetry operation). Thus the fluxes have
apparently a trivial effect, and the disconnectedness of
GL$(N,{\mathbb R})$ is like the factor of 2 in the four connected
components of O$(N,N)$ that likewise had no apparent physical
consequences.

%%%%%%%%%%%%%%%%%%%%%%%%%%%%%%%%%%%%%%%%%%%%%%%%%%%%%%%%%%%%%%%%%%
\section{Super-universality}
\label{sec:superu}

Our results have established universality near the critical point,
at least for the conductance and the mean density states, in the
sense that the universal properties do not depend on the number of
channels, or the details of the disorder, within the symmetry
classes BD and DIII. Similar results were already known for the
orthogonal, unitary, and symplectic chiral symmetry classes
(classes BDI, AIII, and CII, respectively). More surprising is
that the effective FP equation for the transfer matrix near the
critical point, Eq.~(\ref{generalized-eff-theory}), is the same
for all five classes, which implies that the scaling forms for the
distributions both of the conductance and of its logarithm are the
same for all five classes. Moreover, the mean density of states
also has the same form, first found by Dyson\cite{Dyson} for one
case, in all five cases. This suggests that there is
``super-universality'', meaning that {\em all five transitions are
in the same universality class}, even though they lie in distinct
symmetry classes. That is, {\em all} corresponding universal
properties should be the same in the critical region in each
class. It is by now clear that in each symmetry class the
universal properties of the transition can be obtained from the
$N=1$ case. It is of interest to prove the super-universality
directly, and not only for properties related to the transfer
matrix. Here we point out that this can be done within lattice
models for $N=1$; these do {\em not} include random fluxes.

Some results in this direction exist; the equivalence of an $N=1$
model in class BD, which possesses a transition, with one in class
BDI was pointed out in Ref.~\onlinecite{grl1}. This can be
formulated as follows within a 1D tight-binding chain: a
Hamiltonian that consists solely of random imaginary
nearest-neighbor hopping (which lies in class BD with $N=1$) is
equivalent by a simple (disorder-independent) gauge transformation
to a model with real random hopping (which lies in class BDI with
$N=1$).

Now, for each of the three chiral classes, the one-channel
Hamiltonian can be taken to consist solely of nearest-neighbor
hopping that, when hopping to the right on any link, is a random
positive real number (which we call the magnitude of the hopping)
times a random element of ${\mathbb Z}_2$ (i.e.\ $\pm 1$), U($1$)
(i.e.\ a complex number of modulus 1), or SU($2$) (i.e. a
$2\times2$ unitary matrix of determinant 1---in this case only, we
consider particles with spin 1/2), for the orthogonal, unitary,
and symplectic cases, respectively. Clearly, we may perform a
randomness-dependent gauge transformation that eliminates the
group-valued factors and makes the hopping real and positive
everywhere (we assume open, not periodic, boundary conditions).
Then the spectrum of the Hamiltonian (for fixed disorder) is the
same in each class, and independent of the values of the
group-valued disorder. The Green's functions are related by the
gauge transformations. If we assume the magnitude of the hopping
is statistically independent of the group-valued factor on each
link (and the random variables on different links are all
independent), then gauge-invariant moments of Green's functions
are equal for the different classes. These statements are
independent of the distributions for the disorder, apart from the
independence properties already stated. The disorder could be
weak, consisting of small random corrections to a constant (or
more generally, staggered) real positive hopping. This model is
useful in taking the continuum limit to establish contact with
other models in the literature. The disorder could also be taken
with the group-valued factors uniformly distributed over the
relevant group, on each link. In this case, the ``angular''
(group-valued) disorder is strong, though the disorder in the
magnitude could still be weak. In all cases, it follows that all
universal properties at the transitions in these classes are the
same. (It would be surprising if these results were not already
known for the chiral classes.)

This leaves only a mapping for a 1D, $N=1$ model in symmetry class
DIII into either class BD or one of the chiral classes to be
found. It turns out that there is a mapping to the chiral unitary
(AIII) class that parallels that of class BD to the chiral
orthogonal class. We take a tight-binding chain with two basis
states per site. Class DIII Hamiltonians can be considered as
lying in class BD, but with additional restrictions. Thus we may
begin with a purely imaginary (and Hermitian) Hamiltonian on this
chain, and as usual we expect that nearest-neighbor hopping will
be sufficient for our purposes. On examining Altland and Zirnbauer
\cite{AZ}, or Motrunich {\it et al.} \cite{Motrunich} [especially
their equation (5)], we see (further explanation is given below)
that a Hamiltonian in class DIII can be obtained if there are no
on-site terms, and if the hopping matrix between neighbors (going
to the right) takes the form $i$ times a positive real number,
times an orthogonal matrix in O($2$) with negative determinant. By
a disorder-independent gauge transformation, this can be brought
to a real positive number times an SO($2$) matrix. Since this is
equivalent to a complex number, we see that the Hamiltonian is
equivalent to two copies (one the complex conjugate of the other)
of a Hamiltonian in the chiral unitary class. Hence our proof is
complete, and all the universal properties of the transitions in
these five symmetry classes are the same.

We append here a more detailed explanation of the structure of the
$N=1$ class DIII hopping Hamiltonian. Let the components of the
state vector be $y_i$, with $y_{2j-1}$, $y_{2j}$ belonging to the
same site $j$, $j=1$, \ldots, $M$ (in this section, $M$ is a
positive integer). Then the Hamiltonian must be an antisymmetric
imaginary $2M\times 2M$ matrix, i.e.\ an element of the Lie
algebra so$(2M)$ of SO($2M$). By definition \cite{AZ}, a
Hamiltonian in class DIII lies in so$(2M)/$u$(M)$. To understand
the action of u$(M)$, we form the combinations
$z_j=y_{2j-1}+iy_{2j}$. If all $y_i$ are real (an assumption which
does no harm), then the similar combinations
$\overline{z}_j=y_{2j-1}- iy_{2j}$ are simply the complex
conjugates of $z_j$. Now a Hamiltonian that is an $M\times M$
Hermitian matrix acting on the vector of $z_j$ can be viewed as an
antisymmetric matrix acting on the $y_i$. These parametrize a
linear subspace of so$(2M)$, and matrix elements of this form are
supposed {\em not} to be present. That means there are no onsite
terms, and that any non-zero nearest-neighbor terms must violate
the complex structure, effectively mapping $z_j$ to a complex
number times $\overline{z}_{j+1}$. In the $y_i$ components, this
yields the positive real magnitude times $i$ times a $2\times 2$
orthogonal matrix with determinant $-1$, as claimed above. The
nearest-neighbor structure is essential to the equivalence, as the
set of all Hamiltonians in class DIII is certainly not the same as
that for AIII, according to the symmetry classification. However,
there are also Hamiltonians with longer-range hopping that lie in
both classes, and they involve the bipartite lattice structure.

We emphasize that the Hamiltonians here for $N=1$ channel in
classes BD and DIII do not describe superconductors in any simple
direct way, as one finds that $N$ is even in those models. But
they are significant as we have shown earlier that $N=1$ controls
the critical properties in all cases.

The equivalence of the universality classes for the transitions is
not spoiled by the fact that for classes BD and DIII only, one can
include random fluxes that prevent the system from becoming
localized (at $\epsilon=0$). This only affects whether or not the
relevant perturbation that takes the system off the critical point
is possible. The properties at criticality (such as the DOS) are
still the same.

The equivalence of the $N=1$ models considered here suggests that
they may possess additional symmetries, especially when viewed
within the supersymmetry formalism. The corresponding nonlinear
sigma models are certainly not the same, but these correspond to
the $N\to\infty$ limit. Apparently, their $N=1$ analogs must be
equivalent. One example of this phenomenon is known \cite{grl1}.

%%%%%%%%%%%%%%%%%%%%%%%%%%%%%%%%%%%%%%%%%%%%%%%%%%%%%%

\section{Conclusion}
\label{sec:discussion}

We have shown that zero-energy quasiparticle states in a long
disordered superconducting wire with broken spin rotation
invariance and an arbitrary number ($N$) of channels are
generically localized, both in cases with and in cases without
time-reversal invariance. This leads to exponential decay of the
thermal conductance of the wire with its length. Within a model
with disorder invariant under the ``angular'' symmetry group $K$,
there are two localized phases, separated by a critical point
which can be reached by tuning a parameter characterizing the
disorder distribution in the wire. One of our central results is a
universal scaling form for the probability distribution of the
logarithm of the conductance near criticality, Eq.~(\ref{scalG}),
which exhibits two different length scales, which correspond to
the localization lengths for the mean conductance and for the
typical log conductance. We also found the mean density of states
at low energies both at and near criticality, which agree with
Eqs.\ (\ref{DOS-varying}), (\ref{DOS-Dyson}). Our results are in
full agreement with those of Motrunich {\it et al.}
\cite{Motrunich}, who considered the density of states in models
with $N=2$ or $4$ channels. Our results confirm that the universal
results are independent of the number of channels. We showed that
the universality classes for the transitions in the two symmetry
classes are the same, and the same as those in the chiral symmetry
classes. We also showed how the presence of random vector
potentials or fluxes may suppress localization, leaving the system
critical. Our analysis explains why the off-critical behavior was
not detected in some earlier work that used the FP eq.
\cite{BFGM,TBFM}, as the use of a continuum ($a\to0$) limit, or of
$K$-invariant Gaussian disorder, leads to the loss of the relevant
perturbation if $N>2$.

One issue we have left unresolved is the nature of the possible
multicritical point found within some class BD models
\cite{Motrunich}. Another more technical issue is that our
analysis used the generalized FP equation, which was derived on
the assumption that all moments of the increments are finite. It
would be desirable to weaken this assumption as far as possible
(as in the proof of the usual central limit theorem, which
requires only that the second moment of the increments be finite).

We want to emphasize here that our results for the critical regime
are universal, regardless of the number of channels, and thus
regardless of whether the diffusive fixed point is reached as
$L\to0$, which occurs only if $N\to\infty$. The critical (scaling)
regime is defined by the condition that the typical localization
length $\xi$ be much larger than the crossover length $\xi_\times$
from the short length behavior; the critical point corresponds to
$\xi=\infty$. $\xi_\times$ may be of order the mean free path
$\ell$, or much larger $\sim N\ell$ in the many-channel case in
which short-length behavior is diffusive (Ohmic). We used a model
description with disorder invariant under the angular subgroup
$K$. More general models do not have this property at short
scales, but it is expected that the angular variables in the
transfer matrix rapidly become uniformly distributed, so that our
results should be universal for all such models, in the scaling
regime stated.

We also wish to emphasize that our results are generic for classes
BD and DIII. As in the three chiral symmetry classes (BDI, AIII,
and CII), states are generally localized, but there can be phase
transitions as a parameter is tuned. This critical behavior is
intrinsic to these symmetry classes; the behavior must be
described by a parameter that specifies the distance from the
critical point, as well as a mean free path. This may seem to
contradict some widely-held assumptions for 1D disordered wires,
that the symmetry class determines everything, and that the
presence of any other parameter means ``non-universality''.
However, in the general theory of critical phenomena, universal
scaling functions of more than one variable are common. The
necessity of using other parameters in addition to the conductance
at the scale of the mean free path to describe universal phenomena
should be familiar from the example of the quantum Hall effect in
two dimensions, where the additional parameter is the Hall
conductance. Symmetry classes are not universality classes;
universality occurs when the RG flow passes near a fixed point,
and there may be more than just the ballistic, diffusive, and
localized fixed points within a given symmetry class. Which
universality class occurs in a given model within a symmetry class
depends on other aspects of the structure of the model. Examples
of this have been previously observed in class BD in two
dimensions \cite{grl1,Chalk}.

We acknowledge support from the National Science Foundation under
the MRSEC Program, grant no.\ DMR-0213745 (IAG), and under grants
nos.\ DMR-02-42949 (NR) and PHY00-98353 (SV).

%\pagebreak

\appendix

%%%%%%%%%%%%%%%%%%%%%%%%%%%%%%%%%%%%%%%%%%%%%%%%%%%%%%%%%%%%

\section{Invariant differential operators on $G/K$}
\label{app:diff-ops}

In this Appendix we gather necessary information on the spaces of
transfer matrices for classes BD and DIII, viewed as symmetric
spaces $G/K$. A standard detailed reference for analysis on
symmetric spaces and invariant differential operators is
Ref.~\onlinecite{Helgason2}. The appendices in
Ref.~\onlinecite{OP} contain an accessible introduction to root
systems and symmetric spaces.

%%%%%%%%%%%%%%%%%%%%%%%%%%%%%%%%

\subsection{Class BD}

In this case the space of transfer matrices, of type DI in
Cartan's classification, is $M = G/K$, where
\begin{align}
G &= \text{SO}_{0}(N,N), & K &= \text{SO}(N)\times\text{SO}(N).
\end{align}

The Lie algebra of the group $G$ is $\lie(G) =
\text{so}(N,N)$, and it admits the Cartan decomposition:
\begin{align}
\text{so}(N,N) = \text{so}(N) + \text{so}(N) + \cal{P},
\end{align}
where the subspace $\cal{P}$ consists of matrices of the form
\begin{align}
\begin{pmatrix} 0 & \theta \\ \theta^T & 0 \end{pmatrix},
\end{align}
with a real $N\times N$ matrix $\theta$.

Within the space $\cal{P}$ there is a maximal Abelian subspace
$\cal{A}$, and one convenient choice is the set of matrices of the
form
\begin{align}
a_X &= \begin{pmatrix} 0 & X \\ X & 0 \end{pmatrix}, &
X &= \text{diag}(x_1,\ldots,x_N), & x_i &\in \mathbb{R}.
\end{align}
Let us choose a canonical orthonormal basis $\{e_1 \ldots e_N\}$
in the space $\cal{A}^*$ dual to $\cal{A}$ (this space is
isomorphic to ${\mathbb R}^N$ as a vector space) so that
\begin{align}
e_i(a_X) = x_i.
\end{align}
Then the set of roots $\alpha$ of the pair $(\lie(G),\cal{A})$ is
\begin{align}
\Sigma = \bigl\{\pm e_i \pm e_j, \quad 1 \leqslant i < j \leqslant N\bigr\}.
\label{roots}
\end{align}
These vectors form a root system of type D$_N$. All the roots have
the same length, and their multiplicities $m_\alpha$ are all equal
to
\begin{align}
m_o = 1.
\end{align}

It is convenient for our purposes to choose as positive the roots
\begin{align}
\Sigma^+ = \bigl\{e_j \pm e_i, \quad 1 \leqslant i < j \leqslant
N\bigr\}.
\label{roots-positive}
\end{align}
This somewhat non-standard choice leads to the following dominant
Weyl chamber (note that $x_1$ may be positive or negative):
\begin{align}
|x_1| < x_2 < \ldots < x_N.
\end{align}
This makes the middle element in the Cartan decomposition
(\ref{Cartan-decompo-groupbd}) unique.

The half-sum of the positive roots (\ref{roots-positive}) is
\begin{align}
\rho &= \frac{1}{2}\sum_{\alpha \in \Sigma^+} m_\alpha \alpha =
\sum_{i=1}^N (i-1)e_i, \no\\
\rho(x) &\equiv \rho(a_X) = \sum_{i=1}^N (i-1)x_i.
\label{half-sum}
\end{align}
The density function (usually denoted $\delta(x)$ in the
mathematical literature \cite{Helgason2})
\begin{align}
J(x) &= \prod_{\alpha \in \Sigma^+}
\bigl(\sinh \alpha(a_X)\bigr)^{m_\alpha} \no\\
&= \prod_{i<j} \sinh(x_j + x_i) \sinh(x_j - x_i)
\label{density}
\end{align}
is exactly the Jacobian which appears in the Fokker-Planck
equation (\ref{eq:FPbdg}).

The Weyl group $W$ of the root system of type D$_N$ is the
semidirect product of the symmetric group $S_N$ (permuting the
$e_i$) and $\mathbb{Z}_2^{N-1}$ (acting by an even number of sign
changes $e_i \to -e_i$). The ring of polynomial invariants of $W$
is generated by
\begin{align}
s_2, s_4, \ldots, s_{2N-2}, p_N,
\end{align}
where
\begin{align}
s_{2k}(x) &= \sum_{i=1}^N x_i^{2k}, & p_N(x) &= \prod_{i=1}^N x_i.
\label{W-invariants}
\end{align}
We can also form $W$-invariant differential operators on
$\cal{A}$ replacing $x_i$ by $\partial_i \equiv
\partial/\partial x_i$ in these expressions. The
operators obtained in this way,
\begin{align}
s_2(\partial),\ldots, s_{2N-2}(\partial), p_N(\partial)
\end{align}
are local analogs (valid near $x_i=0$ for all $i$) of the
generators
\begin{align}
\Delta_{s_2}, \ldots, \Delta_{s_{2N-2}}, \Delta_{p_N}
\label{radial-parts}
\end{align}
of the algebra of $K$-radial parts of the globally $G$-invariant
differential operators on $G/K$. This may be understood directly,
by comparing these expressions with Eqs.\ (\ref{localDelta-k},
\ref{localDelta-p}), transformed to radial variables, and dropping
the angular parts.

In general, only $\Delta_{s_2} \equiv \Delta_2$, the radial part
of the Laplace-Beltrami operator is known explicitly:
\begin{align}
\Delta_2 = \sum_{i=1}^{N} J^{-1} \frac{\partial}{\partial x_i} J
\frac{\partial}{\partial x_i}.
\end{align}
Conjugation of this expression by the density function leads to
the operator $D_2$ appearing in the FP equation,  and the other
operators $D_{s_{2k}}\equiv D_{2k}$ and $D_{p_N}$ are obtained
similarly; see Eqs.\ (\ref{D2-Delta2}) or (\ref{conjugation}).

For the purposes of this paper we do not need explicit expressions
for the $D$s, but only their asymptotic limits in the regime of a
long wire, when the radial coordinates are widely separated,
$|x_1| \ll x_2 \ll \ldots \ll x_N$. In this regime many
simplifications occur. First, the density function can be
approximated as
\begin{align}
2^{\sum m_\alpha}J(x) &=
e^{\sum m_\alpha \alpha(a_X)}
\bigl(1 + O(e^{-2\alpha_\text{min}(a_X)})\bigr) \no\\
&\approx e^{2\rho(x)} = \prod_i e^{2(i-1)x_i}.
\label{density-asymp}
\end{align}
Secondly, it follows from Theorem II.5.23 in
Ref.~\onlinecite{Helgason2} that the differential operators of
Eq.~(\ref{radial-parts}) tend to
\begin{align}
\Delta_{s_k} \to {\widetilde\Delta}_{s_k} &= e^{-\rho(x)}
s_k(\partial) e^{\rho(x)}
-e^{-\rho(x)} \bigl(s_k(\partial) e^{\rho(x)}\bigr) \no\\
&= \sum_{i=1}^N \left( \bigl(e^{-\rho(x)} \partial_i e^{\rho(x)}
\bigr)^k -\bigl( \partial_i \rho(x)\bigr)^k \right),
\no\\
\Delta_{p_N} \to {\widetilde\Delta}_{p_N} &= e^{-\rho(x)}
p_N(\partial) e^{\rho(x)}
-e^{-\rho(x)} \bigl(p_N(\partial) e^{\rho(x)}\bigr) \no\\
&= \prod_{i=1}^N e^{-\rho(x)} \partial_i e^{\rho(x)}
- \prod_{i=1}^N \partial_i \rho(x).
\label{radial-parts-asymp}
\end{align}
In the subtracted terms the derivatives act only on $e^{\rho(x)}$,
producing $x$-independent constants so that no non-derivative
terms appear in the operators.

Then it follows that the asymptotic forms we need are
\begin{align}
D_{s_k} \to {\widetilde D}_{s_k} &= e^{\rho(x)} s_k(\partial)
e^{-\rho(x)}
-e^{-\rho(x)} \bigl(s_k(\partial) e^{\rho(x)}\bigr) \no\\
&= \sum_{i=1}^N \left( \bigl(e^{\rho(x)} \partial_i e^{-\rho(x)} \bigr)^k
- (i-1)^k \right), \no\\
D_{p_N} \to {\widetilde D}_{p_N} &= e^{\rho(x)} p_N(\partial)
e^{-\rho(x)}
-e^{-\rho(x)} \bigl(p_N(\partial) e^{\rho(x)}\bigr) \no\\
&= \prod_{i=1}^N e^{\rho(x)} \partial_i e^{-\rho(x)}
- \prod_{i=1}^N (i-1).
\label{D-ops-asymp}
\end{align}
For class BD we have (see Eq.~(\ref{half-sum}))
\begin{align}
e^{\rho(x)} \partial_i e^{-\rho(x)} =
\partial_i - i + 1,
\end{align}
so, for example, the FP equation in the asymptotic limit contains
\begin{align}
{\widetilde D}_2 &= \sum_{i=1}^N \bigl( (\partial_i - i + 1)^2
- (i-1)^2 \bigr) \no\\
&= \sum_{i=1}^N \bigl(\partial_i^2 - 2(i-1)\partial_i\bigr).
\label{D2-asymp-BD}
\end{align}
Finally, the operator we are interested in most, $D_{p_N}$, in the
asymptotic limit is replaced by
\begin{align}
{\widetilde D}_{p_N} &= \prod_{i=1}^N (\partial_i - i + 1) =
\partial_1(\partial_2 - 1)\ldots(\partial_N - N +1).
\label{DN-asymp-BD}
\end{align}

%%%%%%%%%%%%%%%%%%%%%%%%%%%%%%%%%%%%%%%%%%%%%%%%%%%%%%%%%%

\subsection{Class DIII}

In this case the space of transfer matrices, of type D in Cartan's
classification, is $M = G/K$, where
\begin{align}
G &= \text{SO}(2N,\mathbb{C}), & K &= \text{SO}(2N).
\end{align}
The Lie algebra $\lie(G) = \text{so}(2N,\mathbb{C})$ of the group
$G$ is usually taken to consist of all antisymmetric complex $2N
\times 2N$ matrices. The Cartan decomposition
\begin{align}
\text{so}(2N,\mathbb{C}) = \text{so}(2N) + \cal{P}
\end{align}
coincides with the decomposition of antisymmetric complex matrices
into real (for $\text{so}(2N)$) and imaginary (for $\cal{P}$)
parts.

By the unitary transformation with
\begin{align}
S = \frac{1}{\sqrt{2}} \begin{pmatrix} I_N & i I_N \\ I_N & -i I_N
\end{pmatrix}
\end{align}
($I_N$ is the $N\times N$ unit matrix) the antisymmetric complex
matrices may be replaced by matrices of the form
\begin{align}
&\begin{pmatrix} A & B \\ C & -A^T \end{pmatrix},
& B^T &= - B, & C^T &= -C,
\label{Lie-algebra-D}
\end{align}
where $A, B, C$ are complex $N\times N$ matrices. In this basis
the Cartan decomposition can be described as follows. Elements of
$\text{so}(2N)$ are matrices (\ref{Lie-algebra-D}) with
antisymmetric real part and symmetric imaginary part, and vice
versa for the elements of $\cal{P}$.

A maximal abelian subspace $\cal{A}$ of $\cal{P}$ is chosen as the
set of matrices
\begin{align}
a_X &= \begin{pmatrix} X & 0 \\ 0 & -X \end{pmatrix}, & X &=
\text{diag}(x_1,\ldots,x_N), & x_i &\in \mathbb{R}.
\end{align}
It follows then that the root system in this case is the same as
for class BD, Eq.~(\ref{roots}), except that the multiplicities
are now
\begin{align}
m_o = 2.
\end{align}

The positive roots and the dominant Weyl chamber are the same as
before, so most of the expressions from the previous subsection
apply here as well. Some changes are
\begin{align}
\rho(x) &= \sum_{i=1}^N 2(i-1)x_i, \\
J(x) &= \prod_{i<j} \sinh^2(x_j + x_i) \sinh^2(x_j - x_i).
\end{align}
The main difference, however, comes from the fact (which is true
whenever the group $G$ is complex) that when all the roots have
multiplicity 2, the function $J^{1/2}(x)$ is a linear combination
of exponentials of positive roots. Then it is possible (see
Theorem II.5.37 in Ref.~\onlinecite{Helgason2}) to find explicit
expressions for all the radial parts of Eq.~(\ref{radial-parts}):
\begin{align}
\Delta_{s_k} &= J^{-1/2} s_k(\partial) J^{1/2}
- J^{-1/2} \bigl( s_k(\partial) J^{1/2}\bigr), \no\\
\Delta_{p_N} &= J^{-1/2} p_N(\partial) J^{1/2}
- J^{-1/2} \bigl( p_N(\partial) J^{1/2}\bigr).
\label{}
\end{align}
As before, the subtractions remove constant non-derivative terms
from the differential operators.

Correspondingly, the generalized Fokker-Planck equation contains
\begin{align}
D_{s_k} &= J^{1/2} s_k(\partial) J^{-1/2}
- J^{-1/2} \bigl( s_k(\partial) J^{1/2}\bigr), \no\\
D_{p_N} &= J^{1/2} p_N(\partial) J^{-1/2}
- J^{-1/2} \bigl( p_N(\partial) J^{1/2}\bigr).
\label{}
\end{align}
As in the previous case of class BD, we do not really need these
explicit expressions, and, instead, use the appropriate asymptotic
limits:
\begin{align}
{\widetilde D}_2 &= \sum_{i=1}^N \bigl(\partial_i^2 -
4(i-1)\partial_i\bigr),
\no \\
{\widetilde D}_{p_N} &= \prod_{i=1}^N (\partial_i - 2i + 2) \no\\
&= \partial_1(\partial_2 - 2)\ldots(\partial_N - 2N + 2).
\label{D2DN-asymp-DIII}
\end{align}

\section{Solution of the Fokker-Planck equation}
\label{app:FP-equation-solution}

Equation (\ref{eq:FPbdg}) is supplemented by the initial condition
which corresponds to ${M} = 1$ for $L = 0$. There are also some
boundary conditions imposed on the function $P(x;L)$. They follow
from the fact that the Fokker-Planck equation (\ref{eq:FPbdg}) has
the form of the continuity equation for the probability current
$S_i$ (we use the short-hand notation $\partial_L =
\partial/\partial L$, $\partial_i =
\partial/\partial x_i$):
\begin{align}
\partial_L P &=
\frac{1}{2 \gamma \ell} \sum_{i=1}^{N}
\partial_i S_i, & S_i = \partial_i P - P \partial_i \ln J.
\end{align}
Then the conservation of the total probability $\int_C dx \,
P(x;L)$, where the integration is over the do\-minant Weyl chamber
$C$, is ensured if the normal components of the current vanish on
the boundary of $C$:
\begin{align}
S_{\hat n}\bigr|_{x \in \partial C} = 0.
\label{boundary-conditions-1}
\end{align}
The boundary $\partial C$ is a union of hyperplanes $x_{i+1}=x_i$
for $i = 1$ through $N-1$, and also $x_2 = - x_1$. Thus we have,
for example,
\begin{align}
\Bigl[\bigl(\partial_i P - P \partial_i \ln J\bigr) -
\bigl(\partial_{i+1} P - P \partial_{i+1} \ln J\bigr)\Bigr]_{x_i =
x_{i+1}} = 0. \label{boundary-conditions-2}
\end{align}
In addition, as we will see momentarily, the function $P(x;L)$
itself also vanishes on $\partial C$, see discussion after
Eq.~(\ref{c-function}):
\begin{align}
P(x;L)\bigr|_{x \in \partial C} = 0. \label{boundary-conditions-P}
\end{align}

The Fokker-Planck equation (\ref{eq:FPbdg}) can be solved in
general (for any symmetry class) with the help of the so called
spherical transform on the space $G/K$, see
Ref.~\onlinecite{Caselle-PRL}. The solution is expressed in terms
of the so called spherical functions $\varphi_k(x)$ labelled by
$N$ real parameters $k = \{k_1,\ldots,k_N\}$. These functions are
the eigenfunctions of $\Delta_2$ with eigenvalue $-k^2 = -\sum_i
k_i^2$. The solution for $P(x;L)$ (up to a $L$-dependent
normalization constant $N(L)$) in terms of the spherical functions
is an analog of the Fourier transform on the double coset
$K\backslash G/K$:
\begin{align}
P(x;L) &= N(L) J(x) \int_{-\infty}^{\infty} \frac{\prod_i
dk_i}{|c(k)|^2} \exp \Bigl(-\frac{Lk^2}{2\gamma\ell} \Bigr)
\varphi_k(x). \label{exact-solution}
\end{align}
The Harish-Chandra function $c(k)$ appearing in the integration
measure is known in all cases, and for the BdG classes is given by
\begin{align}
c(k) &= \prod_{i=1}^{N} \frac{\Gamma\bigl(i\frac{k_i}{2}\bigr)}
{\Gamma\bigl(\frac{m_l}{2} + i\frac{k_i}{2}\bigr)}
%\no\\&\times
\prod_{i<j}^{N}\prod_{\pm} \frac{\Gamma\bigl(i\frac{k_j \pm
k_i}{2}\bigr)} {\Gamma\bigl(\frac{m_o}{2} + i\frac{k_j \pm
k_i}{2}\bigr)}. \label{c-function}
\end{align}
{}From the solution (\ref{exact-solution}) it follows that the
distribution density $P(x;L)$ vanishes on the boundaries of the
Weyl chamber $C$, because the Jacobian $J(x)$ does, and because
the spherical functions $\varphi_k(x)$ are bounded (see Theorem
II.8.1 in Ref.~\onlinecite{Helgason2}).

The spherical functions $\varphi_k(x)$ are known explicitly only
for classes A, AIII, CI, and DIII in the AZ classification, and so
in these cases there are explicit expressions for $P(x;L)$. For
all the symmetry classes, however, the expression
(\ref{exact-solution}) can be significantly simplified in the
asymptotic limit of long wires $L \gg \gamma \ell$. In this case
the ``diffusing'' radial coordinates become very large, $x_i \gg
1$, and for such values of $x_i$ the spherical functions tend to
(see Theorem II.5.5 in Ref.~\onlinecite{Helgason2})
\begin{align}
\varphi_k(x) &\approx J^{-1/2}(x) \sum_{s \in W} c(sk)e^{i(sk,x)},
\label{sperical-function-asymp}
\end{align}
where the summation is over the elements $s$ of the Weyl group
$W$, and $(,)$ denotes the scalar product. It is important that
for classes BD and DIII this is still valid even if $|x_1|$ is not
large. In the limit $L \gg \gamma \ell$, only $k_i \ll 1$
significantly contribute to the integral in
Eq.~(\ref{exact-solution}). For small values of $k_i$ the
Harish-Chandra function becomes
\begin{align}
c^{-1}(k) &\propto \prod_{i=1}^{N} k_i^{\delta(m_l)}
\prod_{i<j}^{N} \bigl(k_j^2 - k_i^2\bigr),
\label{c-function-asymp}
\end{align}
where
\begin{align}
\delta(m_l) &= \biggl\{
\begin{array}{cc}
0, \quad &\text{ for } m_l = 0, \no\\
1, \quad &\text{ for } m_l > 0.
\end{array}
\end{align}
Then the integral in Eq.~(\ref{exact-solution}) can be done, and
with the use of Eq.~(\ref{jacobianJ}) one can obtain
\begin{align}
P(x;L) \approx & N(L) \prod_{i=1}^{N} \exp
\Bigl(-\frac{\gamma\ell}{2L}x_i^2\Bigr) x_i^{\delta(m_l)}
(\sinh 2x_i)^{m_l/2} \no\\
&\times \prod_{i<j}^{N} (x_j^2 - x_i^2) (\sinh^2 x_j - \sinh^2
x_i)^{m_o/2}. \label{asymptotic-solution-1}
\end{align}

Further simplification is possible if we use the fact that $x_i$
($i>1$) are not only large in this regime, but they are also
widely separated:
\begin{align}
|x_1| \ll x_2 \ll \ldots \ll x_N. \label{Weyl-chamber-asymp}
\end{align}
In this portion of the dominant Weyl chamber far from the origin
and the chamber's boundaries the asymptotic solution
(\ref{asymptotic-solution-1}) simplifies further to the form given
in the main text, Eq.~(\ref{asymptotic-solution-2}).

%%%%%%%%%%%%%%%%%%%%%%%%%%%%%%%%%%%%%%%%%%%%%%%%%%%%%%%%%%%%%%%
%%%%%%%%%%%%%%%%%%%%%%%%%%%%%%%%%%%%%%%%%%%%%%%%%%%%%%%%%%%%%%%

\section{Details of the SUSY solution}
\label{app:SUSY}

The supersymmetry technique is most useful for the calculation of
moments of Green's functions. The latter can be viewed as sums
over paths, each path contributing a product of factors that are
the amplitudes (elements of the Hamiltonian or transfer matrix)
for each step of the path. Hence it is natural to represent these
sums in terms of Green's functions in a field theory for particles
(fermions and bosons) that propagate according to these
amplitudes. With equal numbers of fermion and boson components,
the partition function is unity, and the averages of products of
Green's functions can be calculated
\cite{grs,balents-fisher,glr,grl2}. The resulting field theories
possess symmetries (supersymmetry) that rotate fermion and boson
fields into each other. This symmetry guarantees that the
partition function is indeed unity.

In the present case, for class BD, the two directions of
propagation along the wire will be viewed as vertical and labeled
``up'' and ``down'' respectively. The logarithm of the transfer
matrix will effectively become a Hamiltonian that propagates the
particles forwards (up). At each time (i.e.\ position along the
wire, as in the main text), the two sets of variables, up and
down, will be grouped as two ``sites''. We will view this system
then as a two-site quantum Hamiltonian (in imaginary time). For
technical reasons, it is necessary at some stage to treat the
cases $N$ even or odd separately. We consider only $N$ even, as it
describes superconducting wires, and write $N=2n$ for convenience.

On the up site, we may choose the bosons and fermions $b_a, f_a$
(Latin indices run from 1 to $N$, and we assume summation over
repeated indices) to obey canonical commutation relations. On the
down site the bosons $\bb_a$ are canonical, but the fermions
$\fb_a$ satisfy the following commutation relation:
\begin{align}
\{\fbph_a,\fbd_b\} = - \delta_{ab}.
\end{align}
As the dagger $\dagger$ always stands for the adjoint, this
relation results in negative square norms for states with odd
numbers of $\fb$ (if we choose the vacuum to have positive norm).
This is necessary for the cancellations that result in the
partition function being unity (before averaging over disorder).
(Other choices for where the minus signs go are possible, but do
not essentially change any results.) We remark that we could use
$\cal N$ copies of the fermions and bosons (${\cal N}=1$, $2$,
\ldots), and this is necessary when studying higher moments of
Green's function than we will need here.

For the transfer matrices, it is convenient for us here to use the
Cartan decomposition for the transfer matrices at the nodes in the
form
\begin{align}
T = k p. \label{Cartan-T}
\end{align}
The first factor here $k \in K = \text{SO}(N)\times\text{SO}(N)$
describes the propagation and mixing, or forward scattering, of
particle fluxes on the up and down sites (the two factors of
SO$(N)\times$SO$(N)$ correspond to the up and down sites
respectively). As usual, if we assume the uniform distribution for
$k$ with respect to the Haar measure on $K$, then averaging over
this distribution projects the fermionic and bosonic Fock spaces
to the subspaces of the singlets of SO($N$) on the up and
separately on the down sites.

\subsection{The transfer matrix and the Hamiltonian}

The second factor $p \in \exp \cal{P}$ in the Cartan decomposition
(\ref{Cartan-T}) describes the evolution of the states across a
node of the network. The element (\ref{theta}) of $\cal{P}$
parameterized by $\theta$ gives rise to the transfer matrix
\begin{align}
p = \exp \begin{pmatrix} 0 & \theta \\ \theta^T & 0
\end{pmatrix} = \begin{pmatrix} \cosh \sqrt{\theta \theta^T} &
\theta \dfrac{\sinh \sqrt{\theta^T \theta}}{\sqrt{\theta^T \theta}} \\
\theta^T \dfrac{\sinh \sqrt{\theta \theta^T}}{\sqrt{\theta
\theta^T}} & \cosh \sqrt{\theta^T \theta}
\end{pmatrix},
\end{align}
where the matrices $\theta^T \theta$ and $\theta \theta^T$ have
non-negative eigenvalues, and the square roots are well defined.

The second quantized version of the transfer matrix $p$ is the
evolution operator
\begin{align}
V &= \exp{\left(i \theta_{ab} J_{ab}\right)}, \\
J_{ab} &= \bd_a \bbd_b + \bb_b b_a + \fd_a \fbd_b + \fb_b f_a.
\end{align}

As before, we assume that the transfer matrices are random with
some generic $K$-invariant distribution $P(\theta)$. As in our
analysis of the FP equation, it will be sufficient to consider
only the cumulants (\ref{2nd-cumulant}) and (\ref{Nth-cumulant}).
Then, averaging $V$ over the distribution of the disorder
$P(\theta)$ we write it as
\begin{align}
\left[ V \right] = e^{-a H},
\end{align}
where the effective Hamiltonian $H$ contains products of $J_{ab}$
with cumulants of $\theta$, and $a$ is the time step, as before.
The two non-trivial cumulants (\ref{2nd-cumulant},
\ref{Nth-cumulant}) give two terms in the Hamiltonian (we choose
to rescale the Hamiltonian by a factor):
\begin{align}
H_r &= \frac{a}{2c_2} H = H_0 + 2\lambda H_1,  \\
H_0 &= \frac{1}{4} \delta_{a_1 a_2} \delta_{b_1 b_2}
J_{a_1 b_1} J_{a_2 b_2} = \frac{1}{4} J_{ab} J_{ab}, \\
H_1 &= - \frac{N}{2} \frac{i^N}{(N!)^2} \varepsilon_{a_1 \ldots
a_N} \varepsilon_{b_1 \ldots b_N} J_{a_1 b_1} \ldots J_{a_N b_N},
\end{align}
where the parameter
\begin{align}
\lambda = -(N-1)! \frac{c'_N}{2c_2} \label{lambda}
\end{align}
is chosen to simplify equations in the following. Note that this
dimensionless parameter is related to physical length scales
introduced in Section \ref{sec:GenFP} as $|\lambda| =
\xi_\times/(2\xi_\text{typ})$.

As usual, we should add to the Hamiltonian $H_r$ a regulating term
\begin{align}
H_\omega &= \omega (\bd_a\bph_a + \fd_a\fph_a + \bbd_a\bbph_a -
\fbd_a\fbph_a),
\end{align}
which suppresses large boson numbers. $\omega$ represents the
imaginary part of the energy $\epsilon$ in the original random
Hamiltonian, and is necessary when studying e.g.\ the DOS. The
full effective (rescaled) Hamiltonian is then
\begin{align}
H_r = H_0 + 2\lambda H_1 + H_{\omega}. \label{Ham}
\end{align}

\subsection{Supersymmetry algebra \lowercase{osp}($2|2$), and its
representations $R$ and $\Rb$} \label{app:generators-and-states}

The subspaces of states that are invariant under
SO$(N)\times$SO$(N)$ are infinite-dimensional representations of
the superalgebra osp($2|2$), which we will see is the symmetry
algebra of the problem. We denote by $R$ the representation space
on the up site, and by $\Rb$ the one on the down site.

The osp($2|2$) symmetry is generated by the bilinears in the
fermions and bosons that are invariants of SO($N$). The generators
of osp($2|2$) on the up site (in the representation $R$) are given
by (using notation for the generators from Ref.~\onlinecite{snr})
\begin{align}
B &= \frac{1}{2}\left(\fd_a \fph_a - n\right), &
Q_3 &= \frac{1}{2}\left(\bd_a \bph_a + n\right), \no \\
Q_+ &= \frac{1}{2} \bigl(\bd_a\bigr)^2, &
Q_- &= -Q_+^\dagger = -\frac{1}{2} b^2_a, \no \\
V_+ &= \frac{1}{\sqrt{2}} \fd_a \bd_a, &
W_- &= -V_+^\dagger = - \frac{1}{\sqrt{2}} \bph_a \fph_a, \no \\
V_- &= -\frac{1}{\sqrt{2}} \fd_a \bph_a, & W_+ &= -V_-^\dagger =
\frac{1}{\sqrt{2}} \bd_a \fph_a. \label{generators}
\end{align}
Similarly, on the down site (in the representation $\Rb$) the
generators are
\begin{align}
\bar{B} &= \frac{1}{2}\left(\fbd_a \fbph_a + n\right), &
\bar{Q}_3 &= -\frac{1}{2}\left(\bbd_a \bbph_a + n\right),  \no \\
\bar{Q}_+ &= \frac{1}{2} \bb^2_a, &
\bar{Q}_- &= -\bar{Q}_+^\dagger = -\frac{1}{2} \bigl(\bbd_a \bigr)^2, \no \\
\bar{V}_+ &= \frac{1}{\sqrt{2}} \fb_a \bb_a, &
\bar{W}_- &= -\bar{V}_+^\dagger = - \frac{1}{\sqrt{2}} \bbd_a \fbd_a, \no \\
\bar{V}_- &= \frac{1}{\sqrt{2}} \bbd_a \fbph_a, & \bar{W}_+ &=
-\bar{V}_-^\dagger =  -\frac{1}{\sqrt{2}} \fbd_a \bbph_a.
\label{generators-bar}
\end{align}
The generalization to osp$(2{\cal N}|2{\cal N})$ is
straightforward, but not needed here.

\begin{figure*}
\includegraphics[scale=.6]{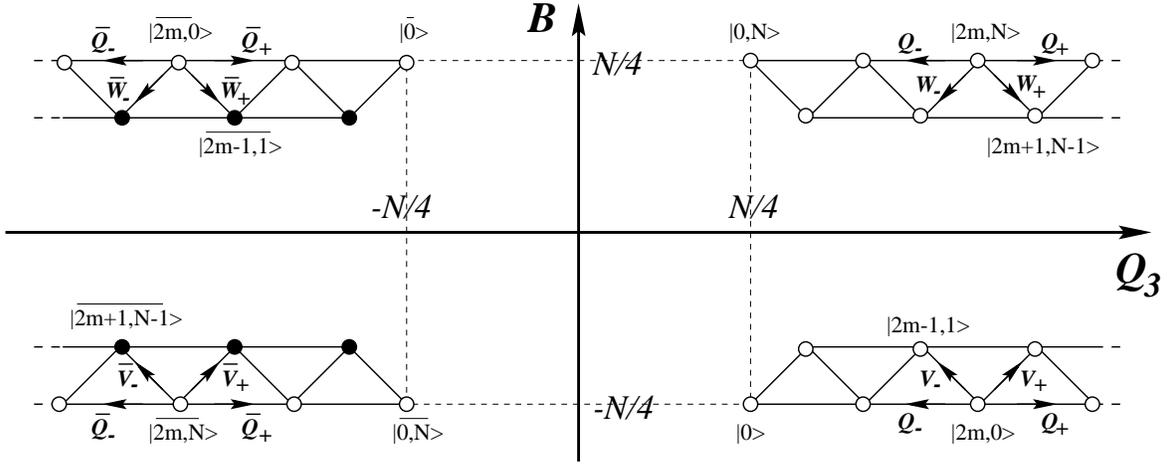}
\caption{Weight diagram for states on the representations $R$ and
$\bar R$. } \label{states}
\end{figure*}

Using these generators we now describe the representations $R$ and
$\Rb$. The representation $R$ splits into two irreducible
representations of the algebra osp($2|2$). The first one of these,
which we denote as $R_0$ is a lowest weight irreducible
representation with the lowest weight state being the vacuum
$|0\ra$ for all $b_a, f_a$. Other states are obtained by repeated
actions of $Q_+$ and $V_\pm$. Normalized states obtained in this
way are
\begin{align}
\st{2m}{0} &=  a_m Q_+^m \sst{0}, \no \\
\st{2m\!-\!1}{1} &=  \bph_m V_- \st{2m}{0},
\end{align}
where
\begin{align}
a_m = \Bigl( \frac{(n-1)!}{m!(m+n-1)!} \Bigr)^{1/2}, \quad b_m =
m^{-1/2}. \label{ambm}
\end{align}

The other subrepresentation of $R$ denoted $R_N$ is also a lowest
weight irreducible representation of osp($2|2$), but now the
lowest weight state is
\begin{align}
\st{0}{N} = \fd_1 \fd_2 \ldots \fd_N \sst{0}.
\end{align}
Other normalized states in $R_N$ are
\begin{align}
\st{2m}{N} &= a_m Q_+^m \st{0}{N}, \no \\
\st{2m\!+\!1}{N\!-\!1} &= c_m W_+ \st{2m}{N}, \label{stdfn}
\end{align}
with $a_m$ as in (\ref{ambm}) and
\begin{align}
c_m = (m+n)^{-1/2}.
\end{align}

The representation $\Rb$ is the dual of $R$. Again, it is split
into $\Rb_0$ and $\Rb_N$ under osp($2|2$). Both these
subrepresentations are highest weight irreducible representations.
In $\Rb_0$ the highest weight state is the vacuum $|{\bar 0}\ra$
for $\bb_a, \fb_a$, and the other normalized states are
\begin{align}
\bst{2m}{0} &= a_m \bar{Q}_-^m \bsst{0}, \no \\
\bst{2m\!-\!1}{1} &= \bph_m \bar{W}_+ \bst{2m}{0},
\end{align}
with the same $a_m, b_m$ as in (\ref{ambm}). Note that the states
with odd number of fermions $\fb_a$ have negative squared norms.
For example,
\begin{align}
\big\langle \overline{2m\!-\!1,1}\bst{2m\!-\!1}{1} = -1.
\end{align}
In $\Rb_N$ the highest weight state is
\begin{align}
\bst{0}{N} = \fbd_1 \fbd_2 \ldots \fbd_N \bsst{0},
\end{align}
and the rest is
\begin{align}
\bst{2m}{N} &= a_m \bar{Q}_-^m \bst{0}{N}, \no \\
\bst{2m\!+\!1}{N\!-\!1} &= c_m \bar{V}_- \bst{2m}{N}.
\end{align}
The states $\bst{2m\!+\!1}{N\!-\!1}$ are negative norm states.

The osp($2|2$) weights of the states in $R$ and $\Rb$ are shown in
Fig.\ \ref{states}. The negative norm states are shown with filled
circles. We also show the action of the osp($2|2$) generators on
the states. We note here that the bosonic part of the supergroup
OSp($2|2$) under consideration is O$(2)\times$Sp$(2,{\mathbb R})$,
and O(2) has two connected components. In the same way that O(2)
has two-dimensional irreducible representations [whereas SO(2),
and its Lie algebra so(2), have one-dimensional irreducibles], the
two subspaces $R_0$ and $R_N$ together form an irreducible
representation of OSp($2|2$) \cite{grl1}. The same is true for the
states in $\Rb$.

The term $J_{ab}^2$ in $H_0$ is an osp$(2|2)$ invariant, and is
proportional to the osp($2|2$)-invariant product of the generators
acting in the representations $R$ and $\bR$. It is convenient to
choose the constant $c_2 = 1/2$, and then
\begin{align}
H_0 &= 2B\bar{B} - 2Q_3\bar{Q}_3 - Q_+\bar{Q}_- - Q_-\bar{Q}_+ \no\\
& \quad - V_+\bar{W}_- + W_-\bar{V}_+ + V_-\bar{W}_+ -
W_+\bar{V}_- . \label{H0}
\end{align}
Then it is clear that $H_0$ acts irreducibly within each of the
four subspaces $R_0 \otimes \bR_0$, \ldots, $R_N \otimes \bR_N$.

The term $H_1$ has a more complicated structure. It is an
osp($2|2$)-invariant product of $N$ operators on each site. In the
representation $R$ we have
\begin{align}
C_+ &= \frac{1}{N!}\varepsilon_{a_1...a_N} \fd_{a_1}... \fd_{a_N},
&
C_- &= C_+^\dagger \no\\
D_+ &= \frac{1}{N!}\varepsilon_{a_1...a_N}
\fph_{a_1}\fd_{a_2}...\fd_{a_N}, &
D_- &= D_+^\dagger \no\\
R_+ &= \frac{1}{N!}\varepsilon_{a_1...a_N}
\bd_{a_1}\bph_{a_2}\fd_{a_3}...\fd_{a_N}, &
R_- &= R_+^\dagger \no\\
X_+ &= \frac{1}{N!}\varepsilon_{a_1...a_N}
\bd_{a_1}\fd_{a_2}...\fd_{a_N}, &
Y_- &= X_+^\dagger \no\\
X_- &= \frac{1}{N!}\varepsilon_{a_1...a_N}
\bph_{a_1}\fd_{a_2}...\fd_{a_N}, & Y_+ &= X_-^\dagger.
\label{nongen}
\end{align}
In the representation $\Rb$ we have
\begin{align}
\bC_- &= \frac{1}{N!}\varepsilon_{a_1...a_N}
\fbd_{a_1}...\fbd_{a_N}, &
\bC_+ &= \bC_-^\dagger \no\\
\bD_- &= \frac{1}{N!}\varepsilon_{a_1...a_N}
\fbph_{a_1}\fbd_{a_2}...\fbd_{a_N}, &
\bD_+ & = \bD_-^\dagger \no\\
\bR_- &= \frac{1}{N!}\varepsilon_{a_1...a_N}
\bbd_{a_1}\bbph_{a_2}\fbd_{a_3}...\fbd_{a_N}, &
\bR_+ &= \bR_-^\dagger \no\\
\bY_- &= \frac{1}{N!}\varepsilon_{a_1...a_N}
\bbd_{a_1}\fbd_{a_2}...\fbd_{a_N}, &
\bX_+ &= \bY_-^\dagger \no\\
\bY_+ &= \frac{1}{N!}\varepsilon_{a_1...a_N}
\bbph_{a_1}\fbd_{a_2}...\fbd_{a_N}, & \bX_- &= \bY_+^\dagger.
\label{nongen-bar}
\end{align}

Then $H_1$ projected onto $R \otimes \bR$ is proportional to the
osp($2|2$) invariant product of these operators, given by
\begin{align}
H_1 &= \frac{N}{2}\bigl[C_+\bC_- + C_-\bC_+ - N(D_+\bD_- + D_-\bD_+) \no\\
& \quad - N(N-1)(R_+\bR_- + R_-\bR_+) \no\\
& \quad - N(X_+\bY_- - Y_-\bX_+ + X_-\bY_+ - Y_+\bX_-) \bigr].
\label{H1}
\end{align}
The terms of $H_1$ that have non-vanishing matrix elements within
$R \otimes \bR$ flip the states between the subspaces $R_0 \otimes
\bR_0$ and $R_N \otimes \bR_N$.

The term
\begin{align}
H_\omega = 2\omega(B + Q_3 - \bar{B} - \bar{Q}_3) \label{Homega}
\end{align}
breaks the supersymmetry from osp($2|2$) down to a u($1|1$)
subalgebra, which is generated by
\begin{align}
Q_3, B, V_-, W_+. \label{u11}
\end{align}

\subsection{Eigenvalue problem}

For non-zero $\omega$ in $H_\omega$ the symmetry of the
Hamiltonian $H_r$ in Eq.~(\ref{Ham}) is u($1|1$) of
Eq.~(\ref{u11}). We can use this to find eigenstates of $H_r$ in
the singlet sector of this u($1|1$). Basis states in this sectors
$\sst{m}_i$ are defined in terms of the states in $R\otimes \Rb$
as follows:
\begin{align}
\sst{0}_1 &= \sst{0}\bsst{0}, \no \\
\sst{m}_1 &= \frac{1}{\sqrt{2}} \Bigl(\st{2m}{0}\bst{2m}{0} \no\\&
+ \st{2m\!-\!1}{1}\bst{2m\!-\!1}{1} \Bigr), & m \geqslant 1, \no \\
\sst{m}_2 &= \frac{1}{\sqrt{2}}
\Bigl(\st{2m\!-\!2}{N}\bst{2m\!-\!2}{N} \no\\& +
\st{2m\!-\!1}{N\!-\!1}\bst{2m\!-\!1}{N\!-\!1} \Bigr), & m
\geqslant 1. \label{singlet}
\end{align}
We look for eigenstates of $H$ of the form
\begin{align}
\sst{\psi} = \psi_0 \sst{0}_1 + \sum_{m=1}^{\infty}(\psi^{(1)}_m
\sst{m}_1 + \psi^{(2)}_m\sst{m}_2), \label{estate}
\end{align}
where the coefficients $\psi_0$, $\psi^{(1)}_m$, and
$\psi^{(2)}_m$ satisfy some recursion relations, and can be found
from them. The relations are obtained by plugging the ansatz
(\ref{estate}) into the eigenvalue problem $H_r \sst{\psi} = E_r
\sst{\psi} \label{evaleq} $ and equating the coefficients of basis
states on both sides of the equation.

First, using the definitions (\ref{generators},
\ref{generators-bar}, \ref{nongen}, \ref{nongen-bar}) of the
operators appearing in the Hamiltonians (\ref{H0}, \ref{H1},
\ref{Homega}), we obtain the action of the parts of the
Hamiltonian $H_r$.
\begin{align}
H_0 \sst{0}_1 &= - n\sqrt{2} \sst{1}_1, \no \\
H_0 \sst{m}_1 &= - (m+1)(m+n)\sst{m+1}_1 \no\\& \quad + m(2m+2n-1)
\sst{m}_1 \no\\& \quad
-(m-1)(m+n-1)\sst{m-1}_1, & m \geqslant 1, \no \\
H_0 \sst{m}_2 &= - m(m+n)\sst{m+1}_2 \no\\& \quad +
(2m-1)(m+n-1)\sst{m}_2 \no\\& \quad - (m-1)(m+n-2)\sst{m-1}_2, & m
\geqslant 1. \\
H_1\sst{0}_1 &=  n \sqrt{2} \sst{1}_2, \no \\
H_1\sst{m}_1 &= (m+n)\sst{m+1}_2 \no\\& \quad
- (m+n-1)\sst{m}_2, & m \geqslant 1, \no \\
H_1\sst{m}_2 &= m\sst{m}_1 - (m-1)\sst{m-1}_1, & m \geqslant
1. \\
H_\omega \sst{m}_1 &= 2m \omega \sst{m}_1, \no\\
H_\omega \sst{m}_2 &= 2(m+n-1)\omega \sst{m}_2, &  \forall m.
\end{align}

Note, that in the subspace spanned by the states $\sst{m}_i$ we
have
\begin{align}
H_1^2 \sst{m}_i = - H_0 \sst{m}_i. \label{H1square}
\end{align}
(It is to have the simplest possible coefficient in this equation
that the constant $\lambda$ was introduced as in
Eq.~(\ref{lambda}).) This relation will significantly simplify the
analysis of the recursion relations for $\psi^{(i)}_m$ for small
$m$ (see section \ref{smallm}).

Combining the above equations, we get the following recursion
relations:
\begin{widetext}
\begin{align}
E_r \psi_0 & = 0,  \no\\
(2\omega - E-R)\psi^{(1)}_1 & = (n+1)\psi^{(1)}_2 -
(2n+1)\psi^{(1)}_1 + n\sqrt{2} \psi_0
+ 2\lambda (\psi^{(2)}_2 - \psi^{(2)}_1), \no \\
(2n\omega - E_r)\psi^{(2)}_1 & = n \bigl(\psi^{(2)}_2 -
\psi^{(2)}_1 + 2\lambda(\psi^{(1)}_1
- \sqrt{2}\psi_0) \bigr), \no \\
-E_r(\Delta \psi^{(1)}_m + \psi^{(1)}_m) & = (m+1) \left(
(m+n+1)\Delta^2\psi^{(1)}_m + (1-2\omega)\Delta \psi^{(1)}_m -2
\omega \psi^{(1)}_m + 2\lambda(\Delta^2 \psi^{(2)}_m +
\Delta \psi^{(2)}_m)\right), & m \geqslant 2, \no \\
-E_r(\Delta\psi^{(2)}_m + \psi^{(2)}_m) & = (m+n) \left(
(m+1)\Delta^2 \psi^{(2)}_m + (1-2\omega)\Delta \psi^{(2)}_m -2
\omega \psi^{(2)}_m + 2\lambda \Delta\psi^{(1)}_m\right), & m
\geqslant 2. \label{diffeqn}
\end{align}
\end{widetext}
The last two equations are written in terms of the finite
differences:
\begin{align}
\Delta \psi_m &= \psi_{m+1} - \psi_m, \no\\
\Delta^2 \psi_m &= \psi_{m+2} - 2 \psi_{m+1} +\psi_m, \label{diff}
\end{align}

Our strategy for the analysis of these equations is the following.
For large $m \gg n$ we can treat $m$ as a continuous variable. The
solutions of the resulting equation are required to decay at
infinity. The solutions of the continuous equation show that the
effect of small $\omega$ becomes significant only for $m \gtrsim
\omega^{-1}$. Then for $m \ll \omega^{- 1}$ we can neglect
$\omega$ in the equation (\ref{diffeqn}). The corresponding
Hamiltonian (within the space $R \otimes \bR$) is simply $2\lambda
H_1 - H_1^2$. Then, instead of solving Eq.~(\ref{diffeqn}), we can
first find the eigenstates of $H_1$ and then combine them to
obtain the energy eigenstates.

After finding approximate solutions in two regions $m \gg n$ and
$m \ll \omega^{-1}$, we can match them in the overlapping region
$n \ll m \ll \omega^{- 1}$ (we choose $\omega$ to be
asymptotically small so that this region is large). For the ground
state this matching procedure gives a complete solution. We can
use this solution to calculate the expectation value of the
``order parameter'' in the ground state, which gives the density
of states in the 1D wire.

For excited states the matching procedure gives the spectrum of
energies $E$ of the coupled superspins. Then the localization
length $\xi_{\text{mean}}$ associated with average correlators in
the state at the Fermi energy in the original 1D wire problem is
given by the inverse of the gap in the energy spectrum above the
ground state in the superspin problem. Let us implement this
strategy step by step.

\subsection{Solution of Eqs.\ (\protect\ref{diffeqn}) for large
\lowercase{$m$}}

For large $m\gg n$, $m$ can be treated as a continuous variable,
and we can replace finite differences by derivatives with respect
to $m$:
\begin{align}
\psi^{(i)}_m \to \psi_i(m), \quad \Delta \psi^{(i)}_m \to
\psi_i'(m), \quad \Delta^2 \psi^{(i)}_m \to \psi_i''(m).
\end{align}
In this limit we neglect $\omega$ and constants of order 1
(including $n$) compared to $m$. Assuming a smooth variation of
$\psi_i(m)$ we also neglect $\psi_i''(m)$ compared to
$\psi_i'(m)$, and $\psi_i'(m)$ compared to $\psi_i(m)$. Then the
Eqs.\ (\ref{diffeqn}) become
\begin{align}
(2\omega m - E_r)\psi_1(m) &= m \bigl(m \psi_1''(m) + \psi_1'(m) +
2\lambda \psi_2'(m)\bigr), \no \\
(2\omega m - E_r)\psi_2(m) &= m\bigl(m \psi_2''(m) + \psi_2'(m) +
2\lambda \psi_1'(m)\bigr). \label{diffleq}
\end{align}

To solve the equtions (\ref{diffleq}), we introduce
\begin{align}
\psi_{\pm}(m)=\psi_1(m) \pm \psi_2(m).
\end{align}
Then, Eqs.\ (\ref{diffleq}) can be written as
\begin{align}
(2\omega m - E_r)\psi_{\pm}(m)  =  m^2 \psi_\pm''(m) + (1 \pm
2\lambda) m \psi_\pm'(m). \label{diffleq2}
\end{align}
After a change of variable $x^2 = 8\omega m$, and the substitution
$\psi_\pm(x)=x^{\mp 2\lambda}f_\pm(x)$, we get the modified Bessel
equation
\begin{align}
x^2 f_\pm''(x) + x f_\pm'(x) - \left(x^2 + 4(\lambda^2 - E_r)
\right) f_\pm(x) = 0.
\end{align}
The decaying solutions for $\psi_{\pm}$ can now be written as
\begin{align}
\psi_{\pm}(x) = 2 A_{\pm} \left(\frac{x}{2}\right)^{\mp 2\lambda}
K_{2\mu}(x), \label{psipmsol1}
\end{align}
where $A_{\pm}$ are constants, $K$ the modified Bessel function,
and
\begin{align}
\mu = \sqrt{\lambda^2 - E_r}. \label{mu}
\end{align}

For the purpose of matching these solutions to the ones for $m \ll
\omega^{-1}$ we need the small $x$ behavior of the Bessel function
$K_{\mu}(x)$. To obtain it we use the definition
\begin{align}
K_{\mu}(x) = \frac{\pi}{2\sin\mu\pi} \bigl(I_{-\mu}(x) -
I_{\mu}(x) \bigr)
\end{align}
and the small $x$ asymptotics for $I_{\mu}(x)$:
\begin{align}
I_{\mu}(x) \sim \frac{(x/2)^{\mu}}{\Gamma(1+\mu)}.
\end{align}
Then we get for small $x$
\begin{align}
K_{\mu}(x) \sim \frac{1}{2}
\left(\Gamma(\mu)\left(\frac{x}{2}\right)^{-\mu} +
\Gamma(-\mu)\left(\frac{x}{2}\right)^{\mu}\right).
\end{align}

After the substitution $x = 2(2\omega m)^{1/2}$ the limiting form
of $\psi_\pm$ for small $m$ becomes
\begin{align}
\psi_{\pm}(m) \sim A_{\pm}\left[\Gamma(2\mu)(2 \omega m)^{-\mu\mp
\lambda} +\Gamma(-2\mu)(2 m \omega)^{\mu \mp \lambda}\right].
\label{psipmsol2}
\end{align}
For the ground state $E_r = 0$, we have $\mu = |\lambda|$, and the
asymptotic solution (\ref{psipmsol2}) takes the form
\begin{align}
\psi_{\pm}(m) \sim A_{\pm}\left[\Gamma(\mp2\lambda) +
\Gamma(\pm2\lambda)(2m\omega)^{\mp 2\lambda}\right].
\label{psigrsol}
\end{align}

Let us specify the previous solutions to the case of $\lambda =
0$. Eq.~(\ref{psipmsol2}) for an excited state wave function
simply becomes
\begin{align}
\psi_\pm(m) \sim A_\pm \left[\Gamma(2\mu)(2 \omega m)^{-\mu} +
\Gamma(-2\mu)(2 \omega m)^\mu\right], \label{asymexc1}
\end{align}
where now $\mu = i \sqrt{E_r}$. For the ground state we have from
Eq.~(\ref{psipmsol1})
\begin{align}
\psi_\pm(m) = 2 A_\pm K_0(x) \sim - A_\pm\bigl(\ln(2\omega m) +
2\gamma\bigr), \label{asymgrst1}
\end{align}
where $\gamma$ is the Euler's constant.

\subsection{Solution of Eqs.\ (\protect\ref{diffeqn}) for small
\lowercase{$m$}} \label{smallm}

In the limit $m\ll\omega^{-1}$, $H_\omega$ can be neglected and
thus we have
\begin{align}
\left[2\lambda H_1 -H_1^2\right]\sst{\psi} = E_r \sst{\psi},
\end{align}
where we used Eq.~(\ref{H1square}). Given the eigenstates of
$H_1$:
\begin{align}
H_1|\psi,j\rangle =  j|\psi,j\rangle, \label{eigH1}
\end{align}
the eigenvalues $j$ are related to $E_r$ by
\begin{align}
j^2 - 2\lambda j + E_r = 0,
\end{align}
which has two solutions
\begin{align}
j_{1,2} = \lambda \pm \mu , \label{j12}
\end{align}
with the same $\mu$ as before, see Eq.~(\ref{mu}). The energy
eigenstates are then given by
\begin{align}
\sst{\psi} = A_1\st{\psi}{j_1} +  A_2\st{\psi}{j_2}.
\label{eneigen}
\end{align}

Equation (\ref{eigH1}) leads to the recursion relations
\begin{align}
j \psi_0 & = 0, \no \\
j \psi^{(2)}_1 & = n(\sqrt{2}\psi_0 - \psi^{(1)}_1),\no \\
j \psi^{(1)}_m & = m (\psi^{(2)}_m-\psi^{(2)}_{m+1}),
& m \geqslant 1, \no \\
j \psi^{(2)}_m & = (m +  n - 1)(\psi^{(1)}_{m-1}-\psi^{(1)}_m). &
m \geqslant 2. \label{recH1}
\end{align}

First, we consider excited states with $E_r > 0$. Then $j \neq 0$,
and it follows from the first of Eqs.\ (\ref{recH1}) that
\begin{align}
\psi_0 = 0.
\end{align}
The rest of Eqs.\ (\ref{recH1}) can be solved using the generating
functions
\begin{align}
f_i(z) &= \sum_{m=1}^{\infty}\psi^{(i)}_m z^{m-1}.
\end{align}
The coefficients $\psi^{(i)}_m$ are obtained from the generating
functions as
\begin{align}
\psi^{(i)}_m &= \oint \!\! \frac{dz}{2\pi i} \, z^{-m}f_i(z),
\label{genfun}
\end{align}
where the integral is taken counterclockwise around zero.

{}From Eqs.\ (\ref{recH1}), we obtain the equations that the
generating functions satisfy:
\begin{align}
z(1-z)\frac{df_1}{dz} + \left[n-(1+n)z\right]f_1 + j f_2 & = 0 \no \\
(1-z) \frac{d f_2}{dz} - f_2 + j f_1 & = 0. \label{geneq}
\end{align}

Substitution of $f_1(z)$ from the second Eq.~(\ref{geneq}) into
the first one, and the change $f_2(z) = (1-z)^{j-1} u(z)$, lead to
a hypergeometric equation
\begin{align}
z(1-z) \frac{d^2 u}{dz^2} + \bigl(n-(n+2j+1)z\bigr) \frac{d u}{dz}
- j(j+n)u = 0,
\end{align}
with the solution which is regular near $z=0$
\begin{align}
u(z) = F(j,j+n;n;z).
\end{align}
Thus, in terms of the hypergeometric function
\begin{align}
\psi^{(2)}_m = A \oint \!\! \frac{dz}{2\pi i} \, z^{-m}(1-z)^{j-1}
F(j,j+n;n;z),
\end{align}
where the constant $A$ is determined later by matching this
solution to the one for large $m$.

To evaluate the last integral, we use one of the Kummer's
relations for the hypergeometric function:
\begin{widetext}
\begin{align}
(1-z)^{j-1} F(j,j+n;n;z) =
\frac{\Gamma(n)\Gamma(-2j)}{\Gamma(n-j)\Gamma(-j)} (1-z)^{j-1}
z^{-j} F(j,1-n+j;1+2j;1-z^{-1}) + (j \leftrightarrow -j).
\end{align}
Then
\begin{align}
\psi^{(2)}_m &=
A\biggl(\frac{\Gamma(n)\Gamma(-2j)}{\Gamma(n-j)\Gamma(-j)}
I(j) + (j \leftrightarrow -j)\biggr), \\
I(j) &= \oint \!\! \frac{dz}{2 \pi i} \, z^{-(m+j)}(1-z)^{j-1}
F(j,1-n+j;1+2j;1-z^{-1}).
\end{align}
In the last integral we deform the contour to run along the branch
cut $[1,\infty)$. On the upper and lower sides of the cut $1-z =
e^{-i\pi}(z-1)$ and $1-z = e^{i\pi}(z-1)$, correspondingly. Then
the sum of the contributions from the two sides is
\begin{align}
I(j) = \frac{\sin \pi j}{\pi} \int_{1}^{\infty}\!\! dz \,
z^{-(m+j)}(z-1)^{j-1} F(j,1-n+j;1+2j;1-z^{-1}). \label{eqI(j)}
\end{align}
The change of variables $x=1-z^{-1}$ leads to a tabulated
integral:
\begin{align}
\int_{0}^{1}\!\! dx \, x^{j-1}(1-x)^{m-1} F(j,1-n+j;1+2j;x) =
\frac{\Gamma(j)\Gamma(m)}{\Gamma(m+j)}\, {}_3
F_2(j,1-n+j,j;1+2j,m+j;1),
\end{align}
\end{widetext}
see Eq.~7.512.5 in Gradshteyn and Ryzhik \cite{gr}. This provides
us with the exact solution of the recursion relations
(\ref{recH1}). However, we really need only the asymptotic form of
this solution for large $m$:
\begin{align}
I(j) = \frac{1}{\Gamma(1-j)}m^{-j}(1+O(1/m)).
\end{align}

Denoting
\begin{align}
\Lambda(j) =
\frac{\Gamma(n)\Gamma(2j)}{\Gamma(n+j)\Gamma(j)\Gamma(1+j)},
%\Gamma(-j)\frac{\sin \pi j}{\pi},
\label{Lambda}
\end{align}
we have the asymptotic behavior of $\psi^{(2)}_m$ for large $m$:
\begin{align}
\psi^{(2)}_m \sim A\bigl(\Lambda(j)m^j + \Lambda(-j)m^{-j}\bigr).
\end{align}
Next we substitute this asymptotic form into the recursion
relation for $\psi^{(1)}_m$ in Eq.~(\ref{recH1}), and obtain with
the same accuracy
\begin{align}
\psi^{(1)}_m \sim - A\bigl(\Lambda(j)m^j -
\Lambda(-j)m^{-j}\bigr).
\end{align}

For a given $j$ we get the asymptotics for the functions
$\psi_\pm$
\begin{align}
\psi_\pm(m) \sim \pm 2A \Lambda(\mp j)m^{\mp j}. \label{psipm-j}
\end{align}
Combining these for both values of $j$, we get, finally, for
excited states
\begin{align}
\psi_\pm(m) \sim \pm 2\bigl(A_1 \Lambda(\mp j_1)m^{\mp j_1} + A_2
\Lambda(\mp j_2)m^{\mp j_2}\bigr). \label{psipm-large-m}
\end{align}

Next, we turn to the ground state. In this case $E_r = 0$, and we
have
\begin{align}
j_1 &= 2\lambda, & j_2 &= 0.
\end{align}
For $j=0$ the Eqs.\ (\ref{recH1}) give
\begin{align}
\psi_0 &= \text{const}, \no\\
\psi^{(1)}_m &= \psi^{(1)}_1 = \sqrt{2}\psi_0, \no\\
\psi^{(2)}_m &= \psi^{(2)}_1.
\end{align}
Then
\begin{align}
\psi_\pm(m) = \sqrt{2}\psi_0 \pm \psi^{(2)}_1.
\end{align}
Combining this with Eq.~(\ref{psipm-j}) we find that the in the
case $\lambda \neq 0$ the Eq.~(\ref{psipm-large-m}) specializes to
\begin{align}
\psi_\pm(m) \sim \pm 2 A_1 \Lambda(\mp 2\lambda)m^{\mp 2\lambda} +
\sqrt{2}\psi_0 \pm \psi^{(2)}_1. \label{limgsol}
\end{align}

In the special case of $\lambda = 0$ the roots $j_1$ and $j_2$
become degenerate, and we have to be more careful. In this case it
is actually easier to come back to the recursion relations
(\ref{diffeqn}). The first of these leaves $\psi_0$ an unspecified
constant, while the rest can now be written as
\begin{align}
(n+1)\bigl(\psi^{(1)}_2 - \psi^{(1)}_1\bigr) &=
n\bigl(\psi^{(1)}_1 - \sqrt{2} \psi_0\bigr),
\no\\
\psi^{(2)}_2 - \psi^{(2)}_1 &= 0, \no\\
\Delta \left[(m+n)\Delta \psi^{(1)}_m\right] &= 0,
\qquad m \geqslant 2, \no \\
\Delta \left[m \Delta \psi^{(2)}_m \right] &= 0, \qquad m
\geqslant 2. \label{zrdfeq1}
\end{align}
Integrating the last pair of equations (\ref{zrdfeq1}), and
finding integration constants from the first pair, we obtain
\begin{align}
(m+n)\left[\psi^{(1)}_{m+1}-\psi^{(1)}_{m}\right] &=
n\bigl(\psi^{(1)}_1 - \sqrt{2} \psi_0\bigr), \no \\
m \left[\psi^{(2)}_{m+1} - \psi^{(2)}_m\right] &= 0.
\label{zrdfeq2}
\end{align}

Thus,
\begin{align}
\psi^{(1)}_{m} &= \psi^{(1)}_{m-1} + \frac{n\bigl(\psi^{(1)}_1 -
\sqrt{2} \psi_0\bigr)}{n+m-1} \no\\
&= \sqrt{2} \psi_0 + n\bigl(\psi^{(1)}_1 - \sqrt{2} \psi_0\bigr)
\left[\Psi(n+m)-\Psi(n)\right], \label{psi1zr}
\end{align}
where $\Psi$ is the Digamma function. For $\psi^{(2)}_m$ on the
other hand, we have
\begin{align}
\psi^{(2)}_m = \psi^{(2)}_1.
\end{align}
For the combinations $\psi_\pm$ this gives
\begin{align}
\psi_{\pm}(m) &= \sqrt{2} \psi_0 \pm \psi^{(2)}_1 \no\\
& + n\bigl(\psi^{(1)}_1 - \sqrt{2} \psi_0\bigr)
\left[\Psi(n+m)-\Psi(n)\right],
\end{align}
with the limiting form, for large $m$,
\begin{align}
\psi_{\pm}(m) &\sim \sqrt{2} \psi_0 \pm \psi^{(2)}_1 \no\\
& + n\bigl(\psi^{(1)}_1 - \sqrt{2} \psi_0\bigr) \left[\ln m -
\Psi(n)\right]. \label{psipmzr}
\end{align}

\subsection{Matching solutions for large and small \lowercase{$m$}}
\label{app:matching}

Let us start with the ground state $E_r = 0$ and a generic
$\lambda \neq 0$. In this case we need to match the Eqs.\
(\ref{psigrsol}) and (\ref{limgsol}). This gives the conditions
\begin{align}
A_{\pm}\Gamma(\mp2\lambda) &= \sqrt{2}\psi_0 \pm \psi_1^{(2)}, \no\\
A_{\pm}\Gamma(\pm2\lambda)(2\omega)^{\mp2\lambda} &= \pm 2 A_1
\Lambda(\mp 2\lambda). \label{match}
\end{align}
Solving for $A_+$ and $A_-$, we get
\begin{align}
A_- &= \frac{2\sqrt{2}\psi_0}
{\Gamma(2\lambda)\bigl(1 - \Phi(\lambda) (2\omega)^{4\lambda}\bigr)}, \no\\
A_+ &= - \frac{2\sqrt{2}\psi_0}{\Gamma(-2\lambda)}
\frac{\Phi(\lambda) (2\omega)^{4\lambda}} {1 - \Phi(\lambda)
(2\omega)^{4\lambda}}, \label{As}
\end{align}
where
\begin{align}
\Phi(\lambda) = \frac{\Gamma^2(-2\lambda)\Lambda(-2\lambda)}
{\Gamma^2(2\lambda)\Lambda(2\lambda)}.
\end{align}
For finite $\lambda$ the leading behavior for asymptotically small
$\omega$ is then given by
\begin{align}
A_+ &\propto \omega^{4\lambda}, & A_- &= \text{const}, &\text{if }
\lambda > 0, \no\\
A_+ &= \text{const}, & A_- &\propto \omega^{4|\lambda|}, &\text{if
} \lambda < 0. \label{Apm}
\end{align}

For $\lambda = 0$, matching Eq.~(\ref{psipmzr}) and
Eq.~(\ref{asymgrst1}) gives
\begin{align}
\psi^{(2)}_m  &=  0, & \forall m, \no \\
A_\pm &= \frac{-\sqrt{2} \psi_0}{\ln(C\omega)}, & C = 2 e^{2\gamma
+ \Psi(n)}. \label{mtchpsi}
\end{align}
Essentially the same result is obtained by taking the limit of
$\lambda \to 0$ in Eqs.\ (\ref{As}). Indeed, in this limit
\begin{align}
& \Gamma(2\lambda) \to \frac{1}{2\lambda}, \qquad
\Phi(\lambda) \to 1, \no\\
& \lim_{\lambda \to 0} \frac{1 - (2\omega)^{4\lambda}}{2\lambda} =
-2 \ln (2\omega),
\end{align}
and Eq.~(\ref{mtchpsi}) follows (but with $C=2$ now).

The final form of the ground state (with the choice of $\psi_0 =
1$) for general $\lambda \neq 0$ is
\begin{align}
\psi_\pm(m) &= A_\pm \Bigl( \Gamma(\mp2\lambda) +
\Gamma(\pm2\lambda)(2m\omega)^{\mp 2\lambda} \Bigr),
& m &\ll \frac{1}{\omega}, \no\\
& = 2 A_\pm (2 \omega m)^{\mp \lambda} K_{2\lambda}(2\sqrt{2
\omega m}), & m &\gg n. \label{gr-state-1}
\end{align}

For $\lambda = 0$ we have, instead,
\begin{align}
\psi_\pm(m) &= \sqrt{2} \biggl( 1 +
\frac{\Psi(m+n)-\Psi(n)}{\ln(C\omega)} \biggr),
& m \ll \frac{1}{\omega} \no \\
&= -\frac{2\sqrt{2}}{\ln(C\omega)} K_0\bigl(2\sqrt{2\omega
m}\bigr), & m \gg n.
\label{gr-state-2}
\end{align}

\begin{figure}[t]
\includegraphics*[width=3.2in]{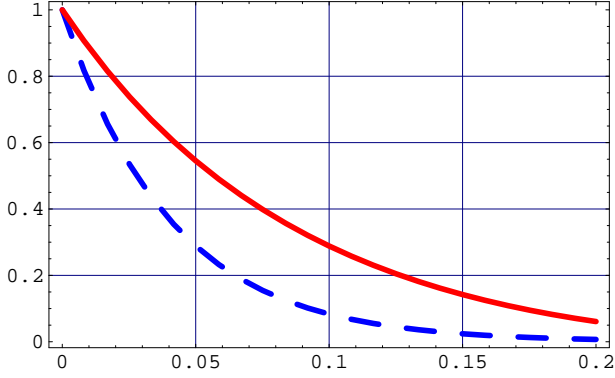}
\caption{Solid and dashed lines show the right and left hand sides
of Eq.~(\ref{omnu}) for $n = 5$, $\lambda = 0.2$, and $\omega =
0.001$. } \label{twoplots}
\end{figure}

Next we discuss excited states with
\begin{align}
E_r = \lambda^2 - \mu^2,
\end{align}
see Eq.~(\ref{mu}). The value of $\mu$ may be obtained by matching
the large $m$ and small $m$ limits of the $\psi^i$'s in the
excited states, Eqs.\ (\ref{psipmsol2}) and (\ref{psipm-large-m})
and by requiring a consistent, non-trivial solution. Unlike in the
ground state, we now have $\psi_0 = 0$, and thus the system of
equations for coefficients $A_{\pm},A_{1,2}$ obtained from the
matching is homogenous:
\begin{align}
A_+ \Gamma(2\mu)(2\omega)^{-\mu-\lambda} &= 2A_1
\Lambda(-\mu-\lambda),
\no\\
A_+ \Gamma(-2\mu)(2\omega)^{\mu-\lambda} &= 2A_2
\Lambda(\mu-\lambda),
\no\\
A_- \Gamma(2\mu)(2\omega)^{-\mu+\lambda} &= -2A_2
\Lambda(-\mu+\lambda),
\no\\
A_- \Gamma(2\mu)(2\omega)^{\mu+\lambda} &= -2A_1
\Lambda(\mu+\lambda).
\end{align}
A non-trivial solution of this system exists only when the
corresponding determinant vanishes, and this gives
\begin{align}
(2 \omega)^{4 \mu} =
\frac{\Gamma^2(2\mu)\Lambda(\mu-\lambda)\Lambda(\mu+\lambda)}
{\Gamma^2(-2\mu)\Lambda(-\mu-\lambda)\Lambda(-\mu+\lambda)}.
\label{omnu}
\end{align}
Before analyzing this equation, let us note that, according to the
definition in Eq.~(\ref{mu}), if $\mu$ is real, then it must lie
in the range
\begin{align}
0 \leqslant \mu \leqslant |\lambda|.
\end{align}
It can also take purely imaginary values.

A quick look at the right hand side (RHS) of the Eq.~(\ref{omnu})
shows that the value $|\lambda| = 1/4$ is special, and we will
restrict ourselves to the values $|\lambda| < 1/4$. In this case
for the allowed real values of $\mu$, the RHS of the
Eq.~(\ref{omnu}) monotonically decreases from 1 to
\begin{align}
-\frac{\Gamma(4|\lambda|)\Gamma(n - 2|\lambda|)}
{\Gamma(-4|\lambda|)\Gamma(n + 2|\lambda|)} > 0. \no
\end{align}
Then for any $n$ there is a value
\begin{align}
\omega_n = \frac{1}{2}\left(-\frac{\Gamma(4|\lambda|) \Gamma(n -
2|\lambda|)} {\Gamma(-4|\lambda|) \Gamma(n +
2|\lambda|)}\right)^{1/4|\lambda|}, \no
\end{align}
such that for $\omega < \omega_n$ there are no real solution for
$\mu$, apart from $\mu = 0$. A typical situation is shown in Fig.
\ref{twoplots}.

The absence of real solutions for $\mu$ means that the energy
spectrum of the reduced Hamiltonian $H_r$ has a gap $\lambda^2$
above the ground state energy (which is 0 by supersymmetry). This,
in turn, implies that the mean localization length in the original
disordered wire problem is finite
\begin{align}
\xi_{\rm mean} = 2 \left(\frac{2c_2}{a} \lambda^2 \right)^{-1},
\end{align}
except at $\lambda =0$, where it diverges. After substitution of
$\lambda$ from Eq. (\ref{lambda}), and $c_2 = a/(\gamma\ell)$, the
last expression reduces exactly to the one given by Eqs.\
(\ref{xitimes}--\ref{kappa'}, \ref{xi-mean}).

For purely imaginary $\mu$, however, Eq.~(\ref{omnu}) has
infinitely many solutions, since $(2\omega)^{2 \mu} = (2\omega)^{2
\mu} e^{2 \pi i l}$, for any integer $l$. Upon expanding both
sides of Eq.~(\ref{omnu}) for small $|\mu|$, one finds that in the
limit of $\omega \ll 1$, $\mu$ takes on a discrete set of values,
\begin{align}
\mu \approx i \frac{\pi l}{2|\ln(2 \omega)|}, \quad l =
0,1,2,\ldots,
\end{align}
which gives for the energy eigenvalues
\begin{align}
E_r = \lambda^2 + \frac{\pi^2 l^2}{4\ln^2 (2\omega)}.
\end{align}
For $\omega \rightarrow 0$, the energy levels above the gap merge
into a continuum.

\subsection{Density of states}

Expressions (\ref{gr-state-1},\ref{gr-state-2}) for the ground
state wave function allow us to determine the leading behavior of
the average density of states $\nu(\epsilon)$ in the wire for
asymptotically low energies $\epsilon$. The density of states
(DOS) $\nu(\epsilon)$ for the original problem of a 1D
superconductor is related to the ``order parameter'' of the
superspin problem. The ``order parameter'' is the measure of the
amount of the symmetry breaking induced by $\omega$ in the ground
state, and is expressed as the ground state expectation value of,
say, the staggered fermionic number
$$
\hat{O} = \bar{B} - B = \frac{1}{2}(2n - n_f - n_\fb).
$$
The density of states is found by analytical continuation $\omega
\to i\epsilon$ as follows:
\begin{align}
\nu(\epsilon) \propto \text{Re} \Bigl( \langle \hat{O} \rangle_0
\bigr|_{\omega \to i\epsilon}\Bigr), \label{dosdef}
\end{align}

Given the form of the singlets in Eq.~(\ref{singlet}), we have
\begin{align}
\hat{O} \sst{0}_1 &= n\sst{0}_1, \no  \\
\hat{O} \sst{m}_1 &= \frac{1}{\sqrt{2}}
\Bigl(n\st{2m}{0} \bst{2m}{0} \no\\
&+ (n-1)\st{2m\!-\!1}{1}\bst{2m\!-\!1}{1}\Bigr), \no \\
\hat{O} \sst{m}_2 &= \frac{1}{\sqrt{2}}
\Bigl((-n\st{2m\!-\!2}{N}\bst{2m\!-\!2}{N} \no \\
&- (n-1)\st{2m\!-\!1}{N\!-\!1}\bst{2m\!-\!1}{N\!-\!1}\Bigr).
\label{actA}
\end{align}
Using the fact that the states $\bst{2m\!-\!1}{1}$ and
$\bst{2m\!-\!1}{N\!-\!1}$ have negative square norms, we obtain
\begin{align}
\langle \hat{O} \rangle_0 &= n\psi_0^2 +
\frac{1}{2}\sum_{m=1}^{\infty} \left[\bigl(\psi^{(1)}_m\bigr)^2 -
\bigl(\psi^{(2)}_m\bigr)^2\right] \no \\
&= n\psi_0^2 + \frac{1}{2}\sum_m \psi_+(m)\psi_-(m)
\label{dosform}
\end{align}

Consider now the case $\lambda = 0$. The sum in
Eq.~(\ref{dosform}) can be broken up into two parts corresponding
to the two regimes, separated by some $m_0$ such that $n \ll m_0
\ll 1/\omega$. The sum over $m \leqslant m_0$ gives a finite
contribution as $\omega \to 0$. The rest of the sum in
Eq.~(\ref{dosform}) dominates for small $\omega$, and can be
approximated by the integral
\begin{align}
& A_+ A_- \int_{m_0}^{\infty}\!\!dm\,
K_0^2(2\sqrt{2 m \omega}) \no\\
&=
\frac{2\psi_0^2}{\omega\ln^2(C\omega)}\int_{x_0}^{\infty}\!\!dx\,
x\, K_0^2(x) \propto \frac{1}{\omega \ln^2(C\omega)}.
\label{m-integral}
\end{align}
Thus, for asymptotically small energy the density of states
behaves as
\begin{align}
\nu(\epsilon) & \sim \text{Re}\frac{1}{i\epsilon \ln^2(C i
\epsilon)} \sim \frac{1}{\epsilon | \ln^3 \epsilon |},
\end{align}
which is precisely the Dyson's singularity of
Eq.~(\ref{DOS-Dyson}).

However, in the case of $\lambda \neq 0$, we have similarly to
Eq.~(\ref{m-integral}) that the contribution from $m > m_0$ to the
``order parameter'' is
\begin{align}
\langle \hat{O} \rangle_0 & \sim A_+A_-
\int^{\infty}_{m_0}\!\!dm\, K_{2\lambda}^2(2 \sqrt{2\omega m})
\sim \omega^{4|\lambda| - 1}.
\end{align}
This contribution dominates the finite contribution from $m < m_0$
only if $|\lambda| \leqslant 1/4$ (for larger values of
$|\lambda|$ our method is not precise enough to determine the
DOS). In this range we then get the following asymptotic form for
the DOS:
\begin{align}
\nu(\epsilon) \sim \epsilon^{4|\lambda| -1}. \label{dos}
\end{align}
This is precisely the power-law with a non-universal exponent
depending on the closeness to criticality, which was found for a
generic 1D system in classes BD and DIII by Motrunich {\it et.
al\/} \cite{Motrunich} in a real space RG study.

\end{document}